\documentclass{emulateapj}

\usepackage{graphicx}
\usepackage[hypertex]{hyperref}

\def \eg {e.g.}
\def \ie {i.e.}
\def\spose#1{\hbox to 0pt{#1\hss}}
\def\ltsim{$\mathrel{\spose{\lower 3pt\hbox{$\sim$}}
        \raise 2.0pt\hbox{$<$}}$\thinspace}
\def\gtsim{$\mathrel{\spose{\lower 3pt\hbox{$\sim$}}
        \raise 2.0pt\hbox{$>$}}$\thinspace}
\newcommand{\thin }{\thinspace}

\newcommand{\lcdm}{$\Lambda$CDM}
\newcommand{\lk }{${\rm L_K}$}
\newcommand{\kms}{${\rm km\ s^{-1}}$}
\newcommand{\li} {${\rm L_I}$}

\newcommand{\msun }{${\rm M_{\odot}}$}
\newcommand{\lsun }{${\rm L_{\odot}}$}
\newcommand{\kfive}{${\rm K_{500}}$}
\newcommand{\msunlsun}{${\rm M_\odot L_\odot^{-1}}$}

\newcommand{\mvir}{${\rm M_{vir}}$}
\newcommand{\rvir}{${\rm R_{vir}}$}
\newcommand{\mfive}{${\rm M_{500}}$}
\newcommand{\rfive}{${\rm R_{500}}$}

\newcommand{\rtwentyfive}{${\rm R_{2500}}$}
\newcommand{\mtwentyfive}{${\rm M_{2500}}$}

\newcommand{\cvir}{${\rm c_{vir}}$}

\newcommand{\reff}{${\rm R_{e}}$}

\newcommand{\src }{NGC\thinspace 1521}
\newcommand{\zfe }{${\rm Z_{Fe}}$}

\newcommand{\survey}{{ElIXr}}
\newcommand{\suzaku}{{\em Suzaku}}
\newcommand{\chandra }{{\em Chandra}}

\newcommand{\xspec }{{\em Xspec}}
\newcommand{\acis }{{\em ACIS}}
\newcommand{\lstar}{${\rm L_*}$}

\newcommand{\ciao }{{\em CIAO}}
\newcommand{\caldb }{{\em Caldb}}
\newcommand{\heasoft }{{\em Heasoft}}
\newcommand{\ned}{{\em{NED}}}

\newcommand{\xmm }{{\em XMM}}

\newcommand{\rosat }{{\em ROSAT}}

\newcommand{\mbh} {${\rm M_{BH}}$}

\newcommand{\lx }{${\rm L_X}$}
\newcommand{\fb}{${\rm f_b}$}
\newcommand{\fbvir}{${\rm f_{b,vir}}$}
\newcommand{\fbtwentyfive}{${\rm f_{b,2500}}$}
\newcommand{\fbfive}{${\rm f_{b,500}}$}

\newcommand{\fgas}{${\rm f_{gas}}$}

\newcommand{\leda}{\href{http://leda.univ-lyon1.fr/}{\em{HyperLEDA}}}

\newcommand{\twomass}{2MASS}
\newcommand{\sigmac}{$\sigma_*$}

\defcitealias{humphrey06a}{H06}
\defcitealias{humphrey08a}{H08}
\defcitealias{humphrey11a}{H11}

\begin{document}
\title{The ELIXR Galaxy Survey. II: Baryons and Dark Matter in an Isolated Elliptical Galaxy}
\author{Philip J. Humphrey\altaffilmark{1}, David A. Buote\altaffilmark{1}, Ewan O'Sullivan\altaffilmark{2,3}, and Trevor J. Ponman\altaffilmark{3}}
\altaffiltext{1}{Department of Physics and Astronomy, University of California at Irvine, 4129 Frederick Reines Hall, Irvine, CA 92697, USA}
\altaffiltext{2}{Harvard-Smithsonian Center for Astrophysics, 60 Garden Street, Cambridge, MA 02138, USA}
\altaffiltext{3}{School of Physics and Astronomy, University of Birmingham, Birmingham, B15 2TT, UK}
\begin{abstract}
The Elliptical Isolated X-ray (\survey) Galaxy Survey is a volume-limited ($<110$Mpc) study of 
optically selected, isolated, \lstar\ elliptical galaxies, to provide
an X-ray census of galaxy-scale (virial mass, \mvir\ltsim $10^{13}$\msun) objects, and 
identify candidates for detailed hydrostatic mass modelling. In this paper,
we present a \chandra\ and \xmm\ study of one such candidate, \src, 
and constrain its distribution of dark and baryonic matter.
We find a morphologically relaxed hot gas halo, extending almost to
\rfive, that is well described by hydrostatic models 
similar to the benchmark, baryonically closed, Milky Way-mass 
elliptical galaxy NGC\thin 720. 
We  obtain good constraints on the enclosed gravitating mass 
($M_{500}$=$[3.8\pm1.0]\times 10^{12}$\msun, slightly higher than NGC\thin 720),
and baryon fraction (\fbfive=$0.13\pm0.03$).
We confirm at $8.2$-$\sigma$ the presence of a dark matter (DM) halo
consistent with \lcdm.
Assuming a Navarro-Frenk-White DM profile, our self-consistent,
physical model enables meaningful constraints beyond \rfive, revealing that most of 
the baryons are in the hot gas. Within the virial radius, \fb\ is consistent
with the Cosmic mean, suggesting that the predicted massive, quasi-hydrostatic gas halos
may be {more common than previously thought}. 
We confirm that the DM and stars conspire to produce an approximately powerlaw 
total mass profile ($\rho_{tot}\propto r^{-\alpha}$) that follows the recently discovered 
scaling relation between $\alpha$ and optical effective radius. {Our conclusions are 
insensitive to modest, observationally motivated, deviations from hydrostatic equilibrium.}
Finally, after correcting for the enclosed gas fraction, 
the entropy profile is close to the self-similar prediction of gravitational structure formation 
simulations, as observed in massive galaxy clusters. 
\end{abstract}



\section{Introduction}
The distribution of mass in galaxies, both in the form 
of baryons and dark matter (DM), is a crucial yardstick for 
elucidating galaxy formation and evolution. 
Dissipationless DM simulations in our current 
(\lcdm) cosmological paradigm predict
DM halos with a characteristic density profile 
\citep{navarro97,navarro04a}, the average shape of which varies
slowly with mass \citep{bullock01a,maccio08a}. In the centres of 
these halos, baryons condense into stars, but the complex interplay
of gas cooling and heating involved (including accretion shocks, feedback
from supernovae, stellar winds, and active galactic nuclei) 
has been the subject of vigorous debate
\citep[\eg][]{white78a,white91a,keres05a,hopkins06a,croton06a}.
Current models tend to predict that large (\gtsim \lstar) galaxies
should possess massive, but diffuse (and, possibly, hard to detect) coronae of 
hot baryons \citep[\eg][]{fukugita06a,crain10a}.

Giant elliptical galaxies  provide a natural laboratory for exploring
these predictions. The 
X-ray emitting hot gas halos around many early-type galaxies 
\citep[\eg][]{osullivan01a} provide a unique opportunity to extend to
a lower mass regime the 
hydrostatic X-ray techniques widely employed to study the dark and 
baryonic matter in galaxy 
clusters \citep[][for a review]{buote11a}.
Provided the hot gas halo is sufficiently bright and morphologically
relaxed, hydrostatic techniques are expected to be useful
\citep{buote95a,rasia06a,nagai07a,piffaretti08a,buote11a}. This is confirmed observationally
by detailed comparisons between masses inferred from 
X-ray methods and independent stellar dynamical measurements, 
or the predictions of stellar population synthesis models, which suggest
a typical accuracy of at least $\sim$20--30\%\ 
\citep[\eg][]{churazov08a,humphrey08a,humphrey09d,shen10a,das10a,das11a,humphrey12b}.


Although the properties
of the largest galaxies  may be intertwined with that of a 
surrounding group or cluster \citep[\eg][]{helsdon01,mathews06a}, 
there is increasing evidence that
$\sim$\lstar\ early-type galaxies in $\sim$Milky Way-mass 
halos can be found with hot gas detectable out at least to
$\sim$\rtwentyfive. In particular, \citet{humphrey06a} presented
a hydrostatic analysis of three isolated galaxies, based on data from
the Chandra X-ray Observatory, inferring virial masses (\mvir)
\ltsim$10^{13}$\msun. Using deep \chandra\ and \suzaku\ data,
\citet{humphrey11a} refined the analysis for one of these
systems, NGC\thin 720,
confirming \mvir=$(3.1\pm0.4)\times 10^{12}$\msun, 
close to the mass of the Milky Way \citep{klypin02a}.
Other X-ray bright, fairly isolated galaxies with similar 
properties have been discussed by \citet{osullivan04b},
\citet{osullivan07b} and \citet{memola11a}.

The extent to which typical giant elliptical galaxies possess
DM halos in accord with \lcdm\ remains unclear.
\citet{buote07a} assembled from the literature 
hydrostatic measurements of the virial mass and concentration
(\cvir=\rvir/$r_s$, where \rvir\ is the virial radius
and $r_s$ is the characteristic scale of the DM halo density
profile) for a sample of galaxies, groups and 
clusters, including the three isolated systems studied by 
\citet{humphrey06a}. They found, for the first time, an
 inverse correlation between \cvir\ and \mvir, with a slope
close to the theoretical value,
confirming a fundamental prediction of \lcdm\ models.
Still, the three isolated galaxies, constituting the low
mass (\ltsim $10^{13}$\msun) end of the relation, were 
marginally ($\sim$2.6-$\sigma$) 
more concentrated than predicted by 
recent theoretical models in the most favourable 
cosmology \citep[the ``WMAP1'' model for relaxed halos reported
by][]{maccio08a}. Given the small number of objects involved,
however, it is unclear whether this represents actual
tension with theory, or is a consequence of selection effects or
small number statistics. 

To date, these represent arguably the best constraints on the 
DM halo concentration for individual giant elliptical galaxies that 
reside in galaxy-scale (\ltsim $10^{13}$\msun) halos. 
Lensing studies alone cannot presently resolve the mass profiles of individual
galaxies, although stacked weak (and strong) lensing analysis
indicates that at least some early-type galaxies can be found in 
$\sim 10^{13}$\msun\ DM haloes with 
concentrations broadly consistent with \lcdm\ \citep{mandelbaum06a,gavazzi07a}.
Orbit- and particle-based stellar dynamical methods are beginning to 
emerge that incorporate DM halos \citep[\eg][]{thomas07b,delorenzo09a},
but there have, so far, been few published constraints on \cvir.

If the DM halos of early-type galaxies are well described by the 
NFW \citep{navarro97} profile, within $\sim$the optical effective
radius (\reff), the baryonic component must be dominant
\citep{buote12a}.
It is increasingly being recognized that the baryons and DM conspire to
produce a {\em total} mass
density profile that 
can be well-approximated by a powerlaw ($\rho_{tot} \propto
r^{-\alpha}$) over a wide radial range 
(\eg\ \citealt{fukazawa06a,gavazzi07a}; \citealt{humphrey10a}, and references therein; \citealt{churazov10a}; \citealt{duffy10a}). 
Although typically $\alpha \simeq 2$ is 
reported \citep{koopmans09a}, \citet{humphrey10a} found that $\alpha$
derived from \chandra\ observations of a
sample of 10 galaxies, groups and clusters
was tightly anti-correlated with \reff. This behaviour
can be understood by the combination of the stellar (Sersic) and DM (NFW)
profiles required to maintain an approximately powerlaw total mass 
distribution, and implies that the DM fraction within \reff\ varies
systematically in such a way as to reproduce, without fine tuning, the 
tilt of the fundamental plane (FP). If this trend is the explanation of the 
tilt of the FP, it should exhibit very little intrinsic scatter, but
more X-ray measurements are needed to investigate this further.
The anti-correlation between $\alpha$
and \reff\ has recently been confirmed by \citet{auger10a}, who used 
joint strong lensing and 
stellar kinematics measurements.  While the optical
results provide important verification of the X-ray work, they are limited
by being confined mostly to within $\sim$\reff, where the DM halo is 
sub-dominant, and because they generally lack 
the resolution to resolve the mass profiles in detail. Indeed, the 
\citeauthor{auger10a} data may be slightly offset from, and exhibit
more intrinsic scatter than, the X-ray relation, although more X-ray data
are needed.

No census of the mass within early-type galaxies is complete
without accounting for the baryons in the hot, diffuse gas.
In the local Universe, the measured stellar and cold
gas content of $\sim$\lstar\ galaxies lies significantly below
the Cosmological baryon fraction \citep[][]{fukugita98a,mcgaugh10a},
and in tension with standard models of galaxy formation 
\citep[\eg][]{benson03a}.
Resolutions to this problem generally involve the ``missing baryons''
either residing in a massive, hot halo \citep[\eg][]{maller04a,fukugita06a} 
or being ejected completely from the system 
\citep[\eg][]{dekel86a,oppenheimer06a}. To date most of the effort to 
locate these hot halos has focused on disk galaxies, where they have not
yet been robustly detected 
\citep[][]{benson00a,anderson10a,rasmussen09a}.
Although X-ray absorption line studies have identified hot gas around the 
Milky Way \citep[\eg][]{nicastro02a,fang02a,rasmussen03a,fang06a,bregman07b,buote09a,fang10a}, and there have 
been reports of extended X-ray emission
unassociated with star formation in a disk galaxy
\citep{anderson11a}, 
whether these constitute the predicted major reservoirs of baryons depends on,
generally uncertain, extrapolation \citep[\eg][]{fang06a,rasmussen09a,anderson10a,anderson11a}.

Recently \citet{humphrey11a} studied the hot gas around the 
isolated, $\sim$Milky Way-mass (\mvir=$[3.1\pm0.4]\times 10^{12}$) 
elliptical galaxy NGC\thin 720, detecting the baryons as far as  
$\sim R_{2500}$. Elliptical galaxies have a distinct advantage over
disk galaxies for detecting the putative hot halos, since few of 
the baryons are bound up in cold gas, so the halo should be denser and
more luminous than in a comparable spiral galaxy. 
In the case of NGC\thin 720, 
the baryon fraction within \rtwentyfive\footnote{We define 
$R_\Delta$ as the three dimensional
radius within which the mean mass density of the system
is $\Delta$ times the critical density
of the Universe.} was tightly 
constrained to \fbtwentyfive$=0.10\pm 0.01$, rising 
to \fb$=0.16\pm 0.04$ by \rvir, consistent with the Cosmological value 
\citep[0.17:][]{dunkley09a}. 
Unlike the ad hoc extrapolations 
often employed (\eg\ the isothermal 
$\beta$-model), this self-consistent evaluation
of the model at large radii required only that the gas is approximately
hydrostatic \citep[as expected around an isolated system,
\eg][]{crain10a}, and the DM mass profile is close to NFW, while 
being relatively insensitive to the thermodynamics of the gas outside
the region where it was clearly detected.
{\em NGC\thin 720 therefore constitutes the most promising detection of a 
baryonically closed
$\sim$Milky Way-mass halo}, {which indicates that feedback need not denude 
such a galaxy of a large fraction of its baryons \citep[\eg][]{kaufmann09a}.
Whether or not NGC\thin 720 represents an unusual case remains to be 
established, but may have clear implications for the location of the 
missing baryons in the local universe.}

Although X-ray observations of galaxy-mass (\ltsim $10^{13}$\msun) halos
provide a unique and powerful insight into galaxy
formation, the small number of reliable measurements
of \cvir, \mvir, baryon fraction (\fb), and $\alpha$ at this mass scale 
 limits the conclusions which can be drawn. 
In addition to the problem of 
small number statistics, the current objects were chosen for study 
heterogeneously, potentially introducing unknown selection effects
\citep[][for a review]{buote11a}.
To address these concerns, we initiated the Elliptical Isolated
X-ray (\survey) Galaxy Survey, an X-ray survey of an optically selected,
volume limited ($<$ 110~Mpc) sample of very isolated, $\sim$\lstar\
early-type galaxies. The isolation condition eliminates most group-scale
halos and ensures an accurate census of the X-ray properties in 
systems resembling the optical properties of 
NGC\thin 720. Isolation also minimizes the likelihood that the low-density
hot gas halo expected around a $\sim$Milky Way-mass galaxy 
is stripped in a dense, cluster environment.
A full description of the sample, and 
initial results are given in \citet{buote12a}. 

Not all of the \survey\ galaxies have luminous X-ray halos
within $\sim$\reff, as might be expected given the large
range of measured  \lx\ at fixed optical luminosity in early-type galaxies
\citep{canizares87a,osullivan01a,ellis06a}. This large scatter
may indicate that both the virial mass of the halo and the feedback
history of the galaxy play a role in determining \lx\
\citep{mathews06a}. Since the gas emissivity depends on the 
square of its density, and gas density profiles are not 
self-similar \citep{humphrey06a,gastaldello07a,sun08a}, it is 
difficult to map galaxies from the  optical {\em versus} X-ray 
luminosity plane onto \fb\ directly. 
This means that \lx, which is typically measured within only a small 
fraction of the virial radius (\rvir), is a poor tracer of the overall 
gas mass \citep[\eg][]{crain10a}. {While even low-\lx\ early-type galaxies
could possess massive hot coronae in which most of the gas has been 
pushed out to large scales, detecting them will pose similar observational
challenges as for spiral galaxies \citep[\eg][]{rasmussen09a}}. Instead, 
those objects with X-ray luminous halos provide arguably the best opportunity
to measure the baryons out to $\sim$tens of kpc, and thus provide
direct constraints on the baryon fraction, 
{at least for a subset of early-type galaxies}. 
It is therefore of interest to define 
an {initial} X-ray luminous sub-sample of the \survey\ galaxies for
further study with detailed hydrostatic methods.

In this paper, we present a detailed \chandra\ and \xmm\ study
of one such galaxy. \src\ was identified 
as one of several X-ray luminous analogues to NGC\thin 720 in a shallow, 
pointed \xmm\ observation taken as part of the \survey\ Galaxy Survey,
and was targetted for deeper follow-up. We discuss the properties of the 
galaxy in detail in \S~\ref{sect_galaxy}, before describing the X-ray 
data-reduction and analysis (\S~\ref{sect_analysis}), the mass modelling
method (\S~\ref{sect_mass}), the likely sources of systematic uncertainty
(\S~\ref{sect_syserr}) and reaching our conclusions in \S~\ref{sect_discussion}.


We adopted a distance of 61.2~Mpc to \src, corresponding to the
redshift 0.01415 \citep{ogando08a}, if we  assume
a flat cosmology with $H_0 =70 {\rm km\ s^{-1}}$
and $\Omega_\Lambda=0.7$. 
In \S~\ref{sect_syserr}, we show that small errors in our distance 
estimate will not affect our conclusions.
At that distance, 1\arcsec\ corresponds to 290~pc. 
We adopted $R_{102}$ as the virial radius (\rvir), 
based on the approximation of \citet{bryan98a} for the redshift of \src.
Unless otherwise stated, all error-bars represent 1-$\sigma$ confidence 
limits (which, for our Bayesian analysis, implies the marginalized region of 
parameter space within which the integrated probability is 68\%).


\section{NGC\thin 1521} \label{sect_galaxy}
\subsection{Target selection}
\begin{figure}
\centering
\includegraphics[width=3.5in]{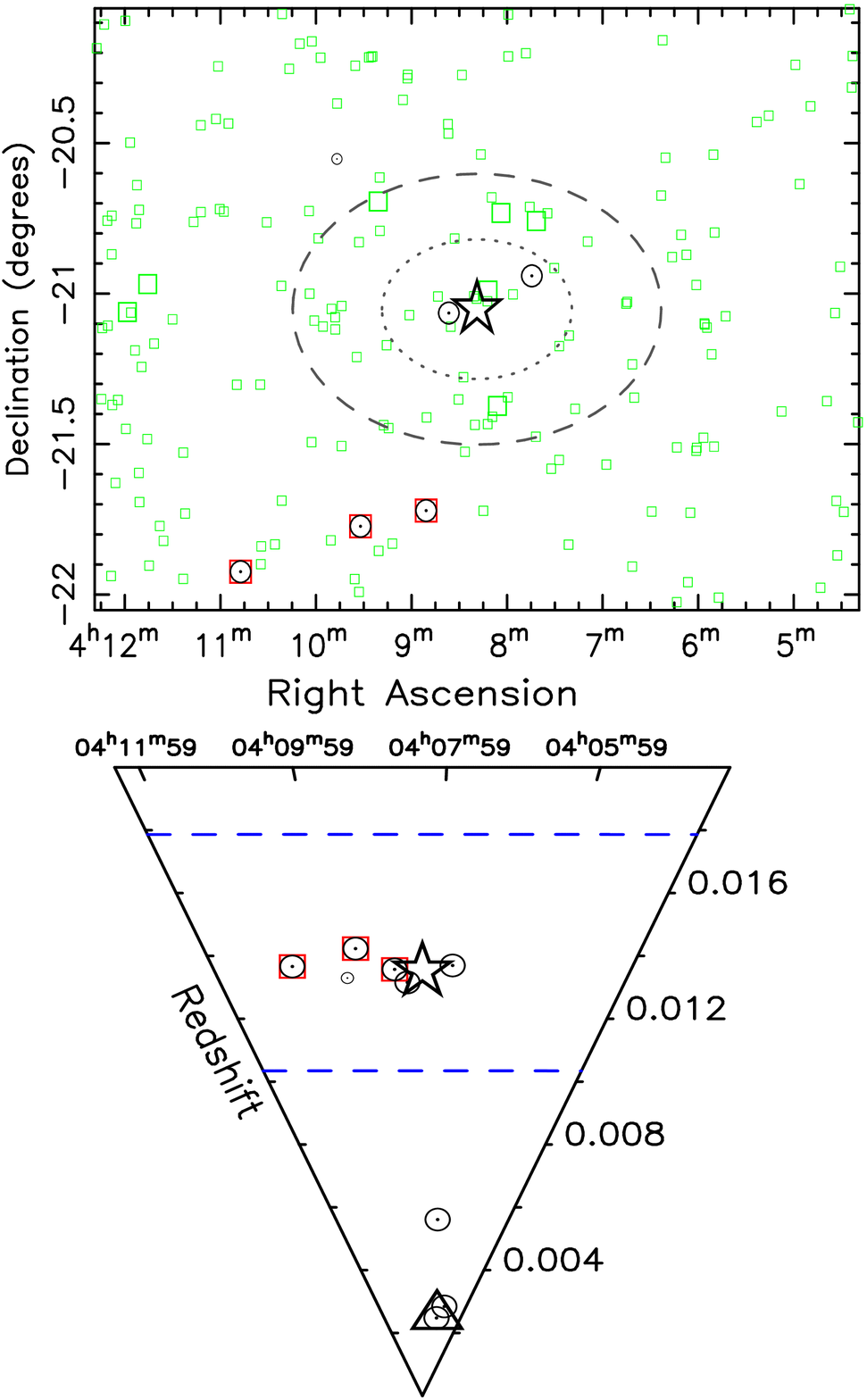}
\caption{Distribution of galaxies around \src\ (shown as a large
star) in the \leda\ database. Galaxies with known redshift are shown as 
large triangles (less than 2 apparent 
magnitudes fainter in the B-band than \src), 
large circles (2--4 magnitudes fainter), or small circles (more than
4 magnitudes fainter). Galaxies with measured recessional
velocities differing by more than 1125\kms\ from \src\
are omitted from the upper panel; none of the galaxies in the upper panel
are within 2 magnitudes of \src.
The corresponding redshift range is shown as dashed blue lines in the lower
plot. Those galaxies in the field of view but without
redshift information are shown as large squares (2--4 magnitudes fainter
than \src\ in the B-band or, if unavailable, in the I-band, K$_s$ or 
``Opt'' band, as given in \leda, respectively), 
or small squares (more than 4 magnitudes fainter). The red squares
indicate the three galaxies catalogued, along with \src, as belonging to the 
small ``group'' S138 by \citet{ramella02a}. The dotted
ellipse indicates the region enclosing projected \rfive, while the dashed 
circle corresponds to \rvir. These regions appear elliptical due to the aspect
ratio of the figure. \src\ is clearly isolated from other 
bright galaxies.} \label{fig_galaxies}
\end{figure}
The \survey\ galaxies were identified in the \leda\ database to be 
optically isolated, $\sim$\lstar\ early-type galaxies within 110~Mpc. 
For a full description of the sample selection and properties,
we refer the reader to \citet{buote12a}. Briefly, we selected early-type 
galaxies 
with absolute B-band magnitude between -21.4 and -19.8, and required
there to be no other galaxy within a projected distance of 750~kpc
that is less than 2 apparent 
magnitudes fainter than the target. Where available,
we used recessional velocity information to eliminate foreground or 
background interlopers (for which we assumed the line of sight velocities
differed from the target by more than 1125\kms). 

In Fig~\ref{fig_galaxies},
we show the distribution of galaxies in the \leda\ database around \src.
Excluding an obvious, bright foreground object (NGC 1518), none of the 
galaxies are less than 2 magnitudes fainter than \src, confirming its
isolation from bright companions. We note that, nevertheless, the group
catalogue of \citet{ramella02a}, derived using a friends-of-friends 
algorithm, included \src\ as a member of the low-mass 
(\mvir$\simeq 10^{13}$\msun) S138 ``group''. The other three 
members of the putative group are marked in 
Fig~\ref{fig_galaxies}, but all are more than 2 magnitudes fainter than
\src\ (so that it does not violate our isolation criterion). Furthermore,
while these galaxies could be gravitationally bound to 
the \src\ system (as the brightest member), they lie outside its virial radius
(\S~\ref{sect_mass}), and are not symmetrically distributed about 
\src, implying that S138 does not represent a virialized
system (whereas \src\ itself does).



\subsection{Optical properties}\label{sect_optical}
\begin{figure}
\centering
\includegraphics[width=3.5in]{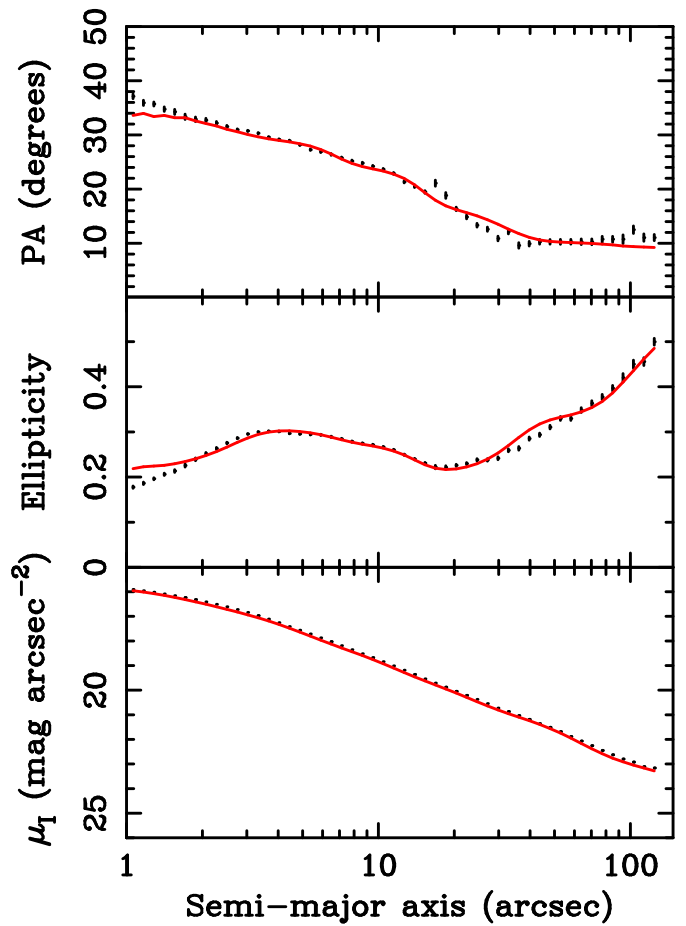}
\caption{{\em Top panel:} I-band isophotal major-axis 
position angle (measured counterclockwise 
from the north) of \src\ (data-points), derived from the 
Carnegie-Irvine Galaxy Survey data \citep{ho11a,li11a}. Overlaid is
 our best-fitting triaxial
model (solid red line). {\em Center panel:} I-band ellipticity profile
and best-fitting model. {\em Lower panel:} I-band major axis 
surface brightness profile of \src, shown along with our best-fitting model. 
The fit is good; the mean absolute residual is only 0.01~mag, which
is sufficient for our purposes. Note that $\mu_I$ differs slightly
from the published profile of \citet{li11a} since we allow the 
isophotal position angle and ellipticity to vary with radius during its 
computation.
\label{fig_optical}}
\end{figure}
{In the optical, \src\ exhibits complex structure.}
B-band isophotal
analysis \citep{capaccioli88a} reveals an ellipticity gradient and
a strong position angle twist within the central $\sim$10~kpc, indicating
that the galaxy is not axisymmetric, and may be triaxial. 
Although our isolation criterion ensures that \src\ is not currently 
undergoing a major galaxy interaction, 
deep B, V, R and I
images taken as part of the Carnegie-Irvine Galaxy Survey 
\citep[CGS:][]{ho11a,li11a}\footnote{\href{http://cgs.obs.carnegiescience.edu/CGS/Home.html}{http://cgs.obs.carnegiescience.edu/CGS/Home.html}} reveal faint 
shell-like surface brightness discontinuities outside the central $\sim$40~kpc,
which may be the relic of a past minor merger event. 
Still, we note that such features may persist in the stars 
for several Gyr following the event
\citep[\eg][]{schweizer92a}, while the sound crossing time of a $\sim$40~kpc 
system, assuming an ambient gas temperature $\sim$0.5~keV/k (similar to what is 
measured in \src; \S~\ref{sect_analysis}), is only $\sim 10^8$~yr. This means that the 
hot ISM will relax to an approximately hydrostatic state following a significant
disturbance far faster than the stellar distribution, and so it 
is unsurprising that the X-ray morphology appears very 
relaxed (\S~\ref{sect_analysis}).

Accurate modelling of the mass distribution requires a reliable deprojection of the 
stellar light profile. To obtain this, we explored a triaxial model for the 
underlying stellar light distribution. Following \citet{cappellari02b}, we 
approximated the stellar density profile as a series of concentric, triaxial ellipsoids
with Gaussian radial profiles ($\rho_* \propto exp\left(-0.5(a_V/\sigma)^2\right)$),
where $\rho_*$ is the stellar density,
$a_V$ is the ellipsoidal coordinate $a_V^2=(x)^2+(y/p)^2+(z/q)^2$ (here $1\ge p\ge q$), 
and x, y and z are Cartesian coordinates. To simplify the 
deprojection, we assumed that the long, short and intermediate axes of each 
ellipsoid were coaligned. For any given orientation (which is 
completely described by three position angles, $\theta$, $\phi$ and $\psi$), 
we projected the ellipsoid onto Cartesian sky coordinates, using 
Eqn~9 of \citet{binney85a}.
We used
dedicated software to fit this model to the central 4.3$\times$4.3\arcmin\
portion of the calibrated, flat-fielded I-band image
produced by the CGS team \citep{ho11a,li11a},
while masking unrelated point sources and other galaxies. To account for seeing,
we convolve the model image by a 1.2\arcsec\ FWHM Gaussian, estimated
from the broadening of stellar images near \src. We adopted the Galactic 
extinction correction from \citet{schlegel98a}.
We allowed the position angles of the system, and the luminosity, $\sigma$, p and q
of each component to vary during the fit. We obtained a satisfactory fit when we 
employed 6 Gaussian components, and not only accurately recovered the 
surface brightness profile (as shown in Fig~\ref{fig_optical}), but also 
the isophotal position angle twist to $\sim 0.5^\circ$ and the ellipticity to $\sim 2$\%, on
average. 



\section{X-ray data reduction and analysis} \label{sect_analysis}
\begin{figure*}
\centering
\includegraphics[width=6in]{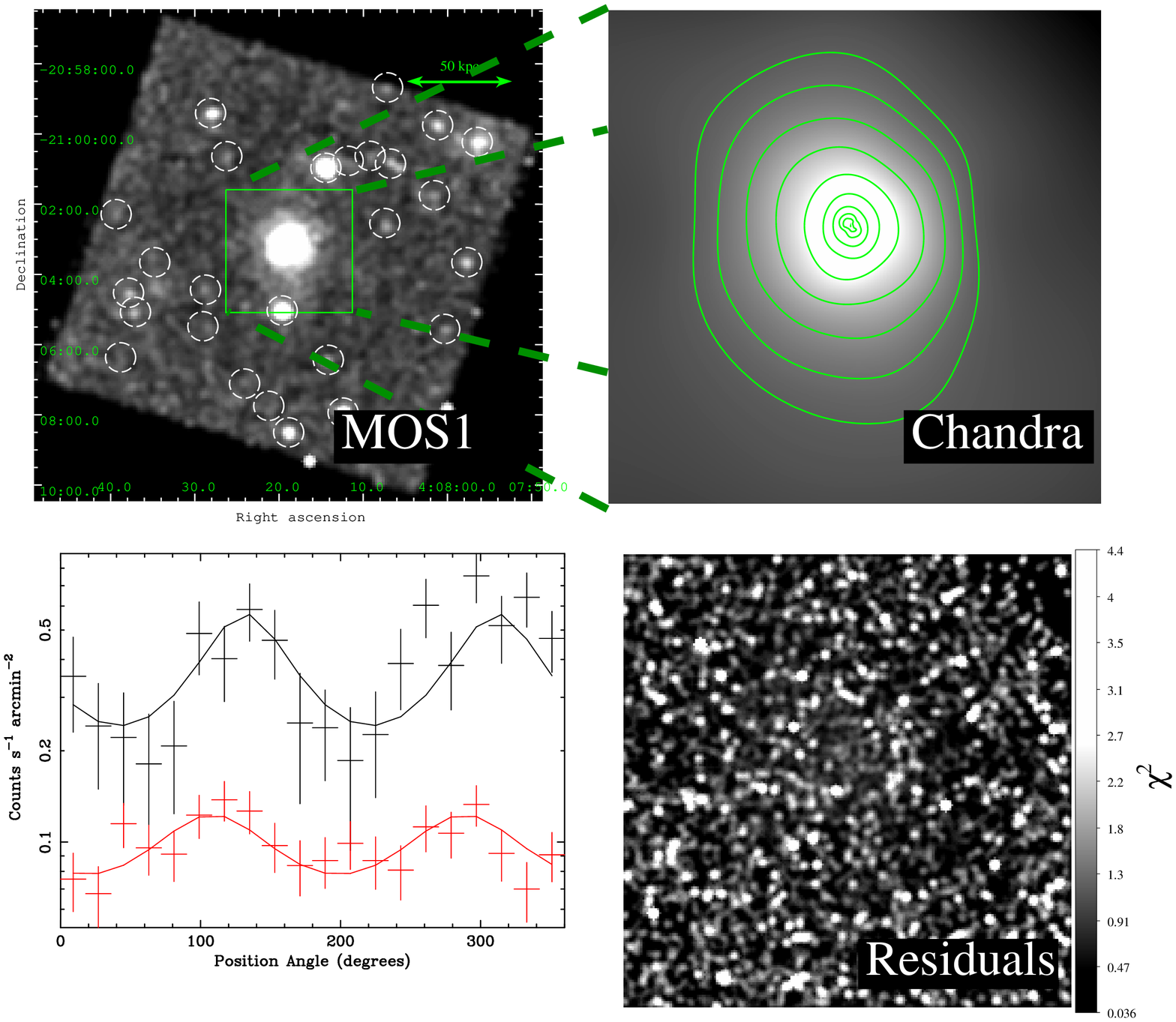}
\caption{\xmm\ MOS1 (central CCD; top left) and \chandra\
(top right) images of \src. Point sources found in the \xmm\
image have been marked with white circles, and are excluded
from subsequent analysis. Although it is not clear given
the dynamic range of the \xmm\ image, the emission from
\src\ extends to the edge of the central CCD, as we show 
in \S~\ref{sect_analysis}. The \chandra\ image has been
cleaned of point sources and mildly smoothed, and arbitrarily spaced
isophotes overlaid to guide the eye. The smoothing scale 
varied from $\sim$1\arcsec\ at the smallest scales, to 
$\sim$0.9\arcmin\ (15~kpc) at the outer part of the image (see text).
To explore the slight distortion of the innermost isophote, we show
(bottom left), the azimuthal variation of the surface brightness
in the (unsmoothed) \chandra\ image, averaged radially between 1--5\arcsec\
(upper, black data-points) and between 5--15\arcsec\ (lower, red
data-points). The position angle is measured anticlockwise from due east. 
We overlay the best 
elliptical $\beta$-model fit (allowing a position angle twist between each
region). We note that the slight excess between
240--300\arcsec\ in the 1--5\arcsec\ data is not statistically
significant. At the bottom right, we show a ``residual significance''
image (see text) of the centre of the system, indicating 
deviations from a smooth fit to the \chandra\ X-ray isophotes. There is no
obvious large-scale feature in this map, indicating the system is 
largely relaxed. \label{fig_images}}
\end{figure*}
\begin{figure*}
\centering
\includegraphics[width=6in]{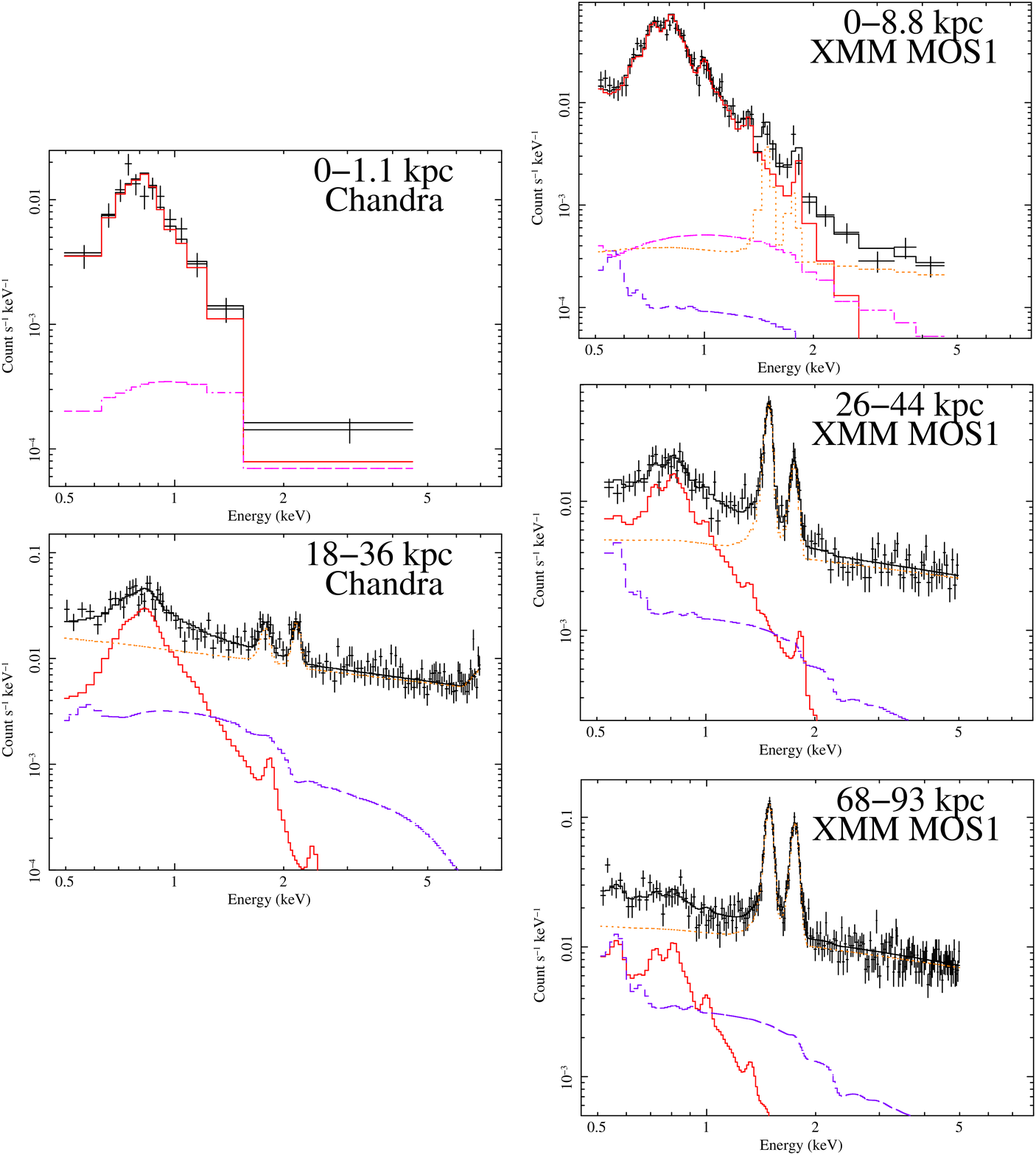}
\caption{Representative \chandra\ and \xmm\ MOS1 spectra for \src, 
shown without background subtraction. In addition to the data, we show the 
best-fitting model, folded through the instrumental response (solid black line), 
along with the decomposition of this model into its various
components. We show the hot gas contribution (sold red line),
the composite emission from X-ray binaries (dash-dot magenta line), 
the instrumental background
(dotted orange) and the cosmic X-ray background (dashed purple line).
The background is dominated by the instrumental component, but
emission from the $\sim$0.5~keV gas is detectable above the 
background below $\sim$1~keV in all the spectra. \label{fig_spectra}}
\end{figure*}
\begin{figure}
\centering
\includegraphics[width=3.4in]{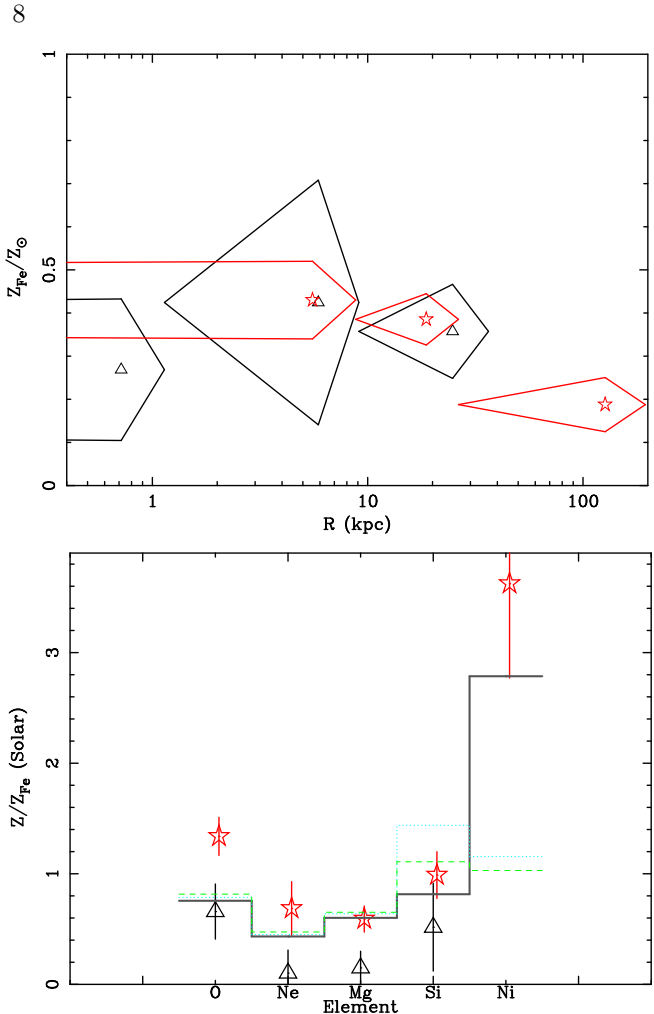}
\caption{{\em Upper panel:} 
Projected Fe abundance profile, measured with \chandra\ (triangles)
and \xmm\ (stars). Note the overall
good agreement between \chandra\ and \xmm. 
{\em Lower panel:} Average abundance ratios with respect to Fe for the 
\chandra\ (triangles) and \xmm\ (stars) data.
The solid line is the best-fit model  where the enrichment comes from SNIa and SNII, assuming
the W7 SNIa yields from \citet{nomoto97b} and the SNII yields from
\citet{nomoto97a}.
The SNIa enrichment fraction is $0.77\pm0.03$. For reference,
we also show the best fits with the WDD1 (green dashed line) and WDD2 
(blue dotted line)
SNIa yields. All abundances are relative to the Solar abundance standard of 
\citet{asplund04a}.\label{fig_abundance}}
\end{figure}
\begin{figure*}
\centering
\includegraphics[width=6in]{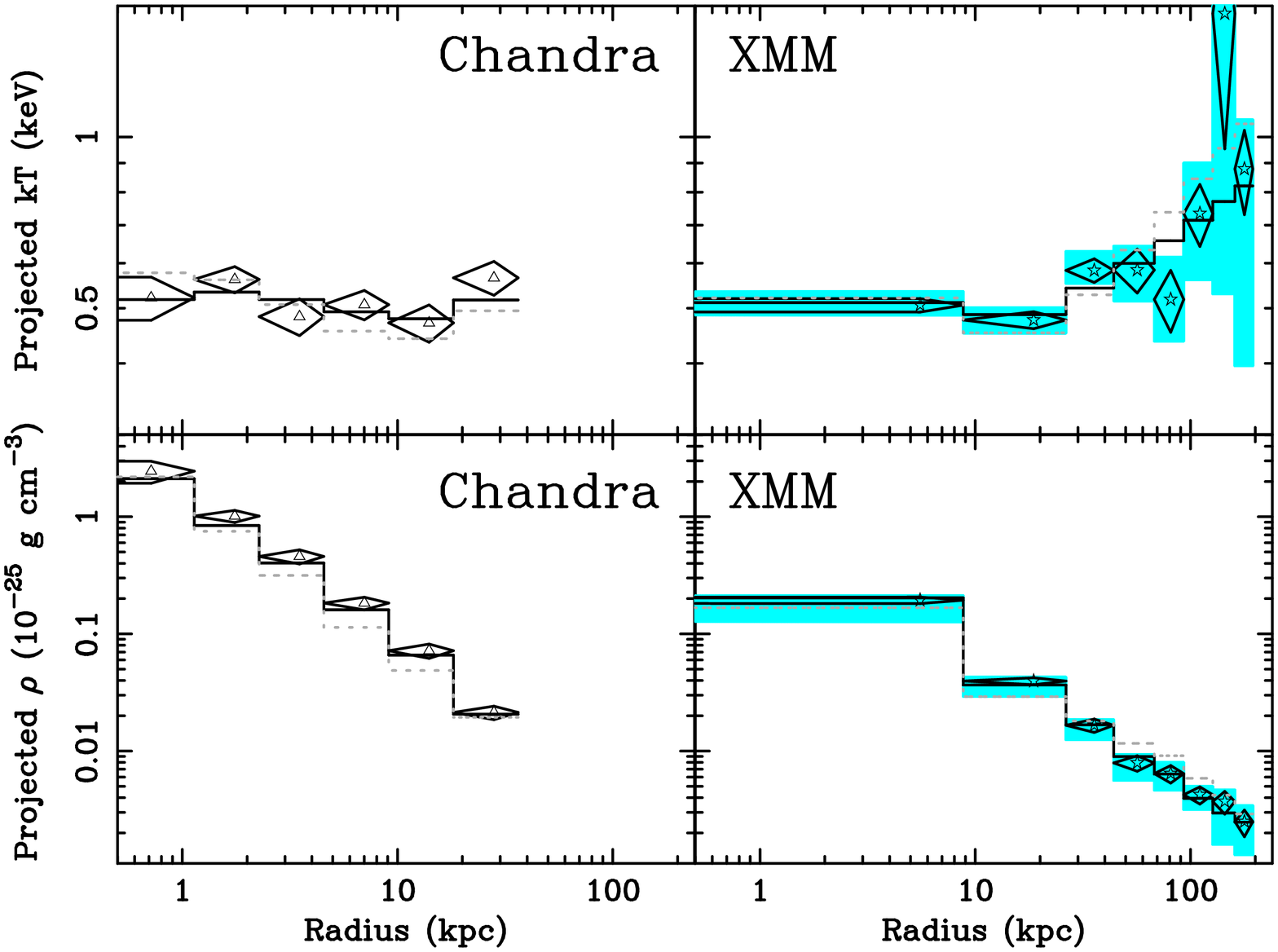}
\caption{Radial temperature (top panels) and density (bottom panels)
profiles for \src. We show the projected \chandra\ profiles in the left
column (triangles), the projected \xmm\ profiles in the right column 
(stars). {The light blue region indicate the effect on the \xmm\ density and temperature profiles of different data-analysis choices; the error-range shown combines both systematic and statistical uncertainties}. Overlaid 
(solid lines) are the 
best hydrostatic model fits to each dataset, which match the data very well.
The dotted lines are the best models if 
dark matter is omitted, for which the fit, in particular to the \xmm\
data, is very poor. 
 \label{fig_kt_rho}}
\end{figure*}
\subsection{\xmm} \label{sect_xmm}
The region of sky containing \src\ was imaged by \xmm\ on 4 separate occasions,
as part of the \survey\ Galaxy Survey. To simplify the analysis,
we consider here only the results from ObsID 0552510101, the deep, 104~ks (excluding periods
of background flaring) exposure beginning on 2009~Feb~7.  In our default analysis, we focus only on the 
EPIC MOS instruments, as we have found the calibration of the PN to be more
uncertain in the low-temperature (kT$\sim$0.5~keV) regime, and the source is 
too faint for the RGS data to be useful. 
Still, as a systematic error check,
we explore the results obtained with the PN in \S~\ref{sect_syserr_instrument}. 
The data were reduced and analysed as described in \citet{gastaldello07a}, using the
XMMSAS 10.0 software suite. Briefly, calibrated events files were generated
with the {\tt emchain} task.  We filtered the list to remove known bright pixels and
hot columns, and periods of high background (flaring) were identified 
visually in the 10--12~keV lightcurve, and excised. 
{Data from the unexposed portions of CCD MOS1-4 exhibited an unusually high count-rate and soft spectrum, consistent with the ``anomalous states'' identified by \citet{kuntz08a}. We therefore excluded all data from this CCD in our subsequent analysis.}
Point sources were
identified by operating on the 0.5--10~keV band images with the {\em ewavelet} task.
The source lists from the three EPIC instruments were merged and checked 
visually. In subsequent analysis, we excluded data from within a 30\arcsec\ 
radius aperture (corresponding to approximately 90\%\ encircled energy)
centred on each confirmed source. 
The image of the central MOS1 CCD
is shown in Fig~\ref{fig_images}, revealing both the extended emission from
\src, and a number of point sources that were excluded.

Spectra were accumulated in 8 concentric, contiguous annuli,
placed at the X-ray centroid, reaching $\sim$190~kpc. 
To mitigate mixing between annuli due 
to the point spread function, the minimum half-width of the annuli (corresponding
to the radius of the central, circular region) was set to 30\arcsec,
corresponding approximately to $\sim$90\%\ encircled energy. Typical spectra
are shown in Fig~\ref{fig_spectra}.
Spectral redistribution matrix files 
(RMFs) and ancillary response files (ARFs) were generated with the SAS 
{\tt rmfgen} and {\tt arfgen} tasks, the latter using an exposure-corrected
detector map to perform flux weighting. 

The MOS1 and MOS2 spectra were fitted simultaneously in \xspec\ vers.\ 12.5.1n, to
obtain the projected abundance, temperature and density profiles
\citep[see][]{humphrey11a}. The data were fitted using the C-statistic,
which is less subject to bias in all count rate regimes than the popular
implementations of $\chi^2$ for Poisson-distributed data 
\citep{humphrey09b}, and the fits were restricted
to the 0.5--5.0~keV band (to avoid instrumental lines at higher energy). 
To aid convergence, we rebinned 
the spectra to ensure at least 20 photons per bin.
We modelled the hot gas as an APEC model,
and included a 7.3~keV bremsstrahlung component to account for undetected
LMXBs, which was only significant within $\sim$\reff\ 
{\citep[4.7~kpc, from the \twomass\ database:][]{jarrett00a}}.
Hot gas abundance ratios with respect to Fe were tied between
all annuli. Where they could not be constrained, they were fixed at the 
Solar ratio \citep{asplund04a}.

{To account for the background, we 
adopted the approach in \citet{humphrey11a}. Specifically, to 
account for the instrumental and particle background, we included a broken powerlaw
model (not folded through the ARF) and two Gaussians 
(at 1.5~keV and 1.7~keV). The normalization of the instrumental components was allowed to vary with 
radius, and (to improve constraints) the shape of the broken powerlaw component was tied between the 
inner three annuli, and between the fourth and fifth annuli. 
We note that \citet{kuntz08a,snowden08a} provide an alternative strategy for background subtraction (ESAS\footnote{\href{http://heasarc.gsfc.nasa.gov/docs/xmm/xmmhp_xmmesas.html}{http://heasarc.gsfc.nasa.gov/docs/xmm/xmmhp\_xmmesas.html}}),
which we explore in detail in \S~\ref{sect_syserr_background}. We choose not to use 
ESAS for our default analysis since there are inherent uncertainties in the
procedure to map from the out of field of view count rates onto the instrumental background, 
which could cause problems in the highly background-dominated regime at the outskirts of \src.
To account for the cosmic X-ray background, 
we included an (absorbed) powerlaw model with $\Gamma$=1.41 \citep{deluca04a}.

Given its Galactic coordinates (l=$216^\circ$, b=$-45^\circ$), 
\src\ is located at the extreme edge of the larger of two adjacent
excesses in the soft X-ray background, known as the ``Eridanus X-ray 
enhancement'', that may be due to an old supernova remnant 
\citep{naranan76a,snowden95a}. Therefore, it is important to take care
over the characterization of the soft X-ray background. Using pointed
\xmm\ observations,  \citet{henley10a} were able to characterize the 
Galactic background at high latitudes, including data 
in the vicinity of \src, with 
a model comprising two plasma components (one unabsorbed and one absorbed), 
with kT$\sim$0.1~keV and $\sim$0.2~keV, respectively \citep[see also][]{kuntz00a}. 
For the two pointings within
$\sim$8$^\circ$ of \src\ (their observations 22 and 23), kT
of the hotter plasma component was constrained to $0.197^{+0.012}_{-0.018}$~keV
and $0.189^{+0.009}_{-0.014}$~keV, respectively, which are consistent within errors. 
The \rosat\ 0.1--1.0~keV count-rate in a 0.6--1.0$^\circ$ region around 
\src\footnote{Determined by querying the \rosat\ diffuse X-ray
background maps \citep{snowden97a} with the on-line HEASARC X-ray background tool,
\href{http://heasarc.gsfc.nasa.gov/cgi-bin/Tools/xraybg/xraybg.pl}{http://heasarc.gsfc.nasa.gov/cgi-bin/Tools/xraybg/xraybg.pl}} is actually within $\sim$15\%\ of that coincident with
 \citeauthor{henley10a}'s observation 23. 

Since \src\ is closer to the peak 
of the Eridanus enhancement than these fields, the temperature of the 
hotter component could, plausibly, be different. To explore this, we 
analysed a deep ($\sim$100~ks) archival \suzaku\
observation of the ``Eridanus Hole'' (ObsID 502076010), a blank-sky
region also in the outskirts of the Eridanus enhancement (l=$213^\circ$, b=$-39^\circ$), 
but where the background is 
$\sim$30\%\ higher than for \src. We reduced the data as described in \citet{humphrey11a},
and extracted a spectrum from the whole field, excluding data in 2\arcmin\ regions around
bright point sources (we identified these sources by eye in the \xmm\ EPIC MOS1 
image of the same field (ObsID 0203900101); we used \suzaku, rather than \xmm\ for characterizing the 
Galactic emission due to its lower, stabler instrumental background and good spectral resolution). 
As this is effectively a blank-sky field, after subtracting off the standard instrumental
background (which was estimated by standard \heasoft\ tools; for more details, refer to \citealt{humphrey11a}), 
the remaining spectrum represents the 
X-ray background. We were able to fit this in the 0.5--5.0~keV band with a model comprising a
$\Gamma=1.41$ powerlaw, an unabsorbed $0.07$~keV APEC plasma model \citep[with Solar abundances:][]{asplund04a}, 
and an absorbed APEC model with kT=$0.22\pm0.02$~keV, \ie\ approximately the same model as was used by 
\citet{henley10a}. Although the inferred distance of the gas contributing to the Eridanus enhancement
is close \citep[$\sim$400~pc; \eg][]{snowden95a}, we found an adequate fit using the whole
Galactic column for the absorbed APEC component, which may reflect the high galactic latitude
of the feature. Freeing $N_H$ did not give rise to a significantly
better fit. We therefore adopted kT of 0.07 and 0.20~keV, respectively for the two APEC components in
our fit, and used the whole Galactic column density \citep{kalberla05a} to absorb the hotter
component.}



We found this background model was able to fit adequately the background 
spectra generated for regions corresponding to each annulus used in our 
\xmm\ analysis from the standard ``template''
events files. We found that our full model was able to fit the spectra in 
all annuli reasonably well, as shown in Fig~\ref{fig_spectra}.
We show the inferred Fe abundance profile
and the global abundance ratios in Fig~\ref{fig_abundance}, and 
the measured temperature and density profiles  in Fig~\ref{fig_kt_rho}.
We note that we were able to obtain good constraints on the gas 
temperature and density out to 190~kpc, which is well beyond
\rtwentyfive, and is approaching \rfive\ (\rtwentyfive=$121\pm8$~kpc; \rfive=$240\pm22$~kpc; \S~\ref{sect_mass}), which is beyond even what 
we achieved in our NGC\thin 720 analysis.
{We expect the results not to be strongly sensitive to the treatment of the background
if the ratio of the source to the background rate \gtsim 0.5 in the 0.65--0.9~keV band, corresponding 
to the Fe L-shell region, which is the crucial temperature diagnostic for a $\sim$0.5~keV
plasma. This is true for the inner 5 \xmm\ annuli, but for the outer 3 annuli, the 
source/ background ratio falls to $\sim$0.22, 0.08 and 0.07, respectively. 
We explore the impact of different background subtraction subtraction methods in detail
in \S~\ref{sect_syserr_background}, finding, unsurprisingly that the outermost annuli
are most sensitive to this choice, as shown in Fig~\ref{fig_kt_rho}. Nevertheless, we 
do not reach qualitatively different conclusions regarding the global 
parameters of the system (\S~\ref{sect_syserr_background})}.


\subsection{Chandra}
The region of sky containing \src\ was imaged by the \acis\
instrument aboard \chandra\ (ObsID 10539; beginning on 2009 Jul 4) 
for a total of 49~ks good time (with periods of background flaring removed).
The data were reduced and analysed as described in \citet{humphrey12a},
using the \ciao\ 4.3 software suite and
the corresponding \chandra\ calibration database (\caldb) vers.\ 4.4.2.
Briefly, the data were reprocessed from the ``level 1'' events files, following the
standard data reduction threads\footnote{http://cxc.harvard.edu/ciao/threads/index.html}. 
Periods of high background were identified by eye in the lightcurve from a low surface-brightness
region of the CCDs and data from these intervals were excised. Point sources were 
detected in the 0.3--7.0~keV image with the {\tt wavdetect} \ciao\ task. All sources 
were confirmed visually, and appropriate elliptical regions containing 
$\sim99$\%\ of the source photons were generated. Data from these regions were excluded in 
subsequent analysis.

In Fig~\ref{fig_images} (top right), we show a 
smoothed, flat-fielded \chandra\ image, having removed the point sources with the 
algorithm outlined in \citet{fang09a}. The image was smoothed with a Gaussian kernel, the 
width of which varied with distance from the nominal X-ray centroid 
according to an arbitrary powerlaw, ranging from $\sim$1\arcsec\ at the centre of the 
image to $\sim$0.9\arcmin\ at its edge. The image is smooth, albeit slightly elliptical.
To search for more subtle structure
we used dedicated software to fit an elliptical beta model (with constant ellipticity)
to the unsmoothed (flat-fielded) image.
In Fig~\ref{fig_images} (bottom right), we plot $(data - model)^2/model$, corresponding 
(approximately) to the $\chi^2$ residuals from this fit. To bring out the structure,
we smoothed this image with a Gaussian kernel of width 3~pixels. 
The lack of significant, coherent residuals indicates that the X-ray image is very relaxed.
An alternative view of the central (\ltsim 15\arcsec) region is given in the lower left
panel, which shows the azimuthal variation in the surface brightness, integrated over
two radial bins. There are no statistically significant residuals from an elliptical
$\beta$-model fit.
{Although the central isophotes in the smoothed \chandra\ image do 
appear slightly more flattened, this feature is not statistically significant, and may 
represent a statistical fluctuation. }

Spectra and associated flux-weighted responses were extracted in a series of 
contiguous, concentric annuli placed at the 
X-ray centroid. The widths of the annuli were chosen to contain approximately the 
same number of background-subtracted counts, while ensuring sufficient photons for
useful spectral analysis. The resulting annuli had widths larger than $\sim$3\arcsec,
which is sufficient to prevent spectral mixing between adjacent annuli on account
of the finite spatial resolution of the mirrors. Data in the vicinity of point sources
and chip gaps were excluded. The spectra were fitted simultaneously with \xspec,
similarly to the \xmm\ data, and the fit was restricted to the 0.5--7.0~keV band. To
model the background, we used a similar approach to our \xmm\ analysis, although the 
surface brightness of the background components was assumed to be constant over the region
of interest (which is entirely confined to the S3 chip). 
Representative spectra, and the best-fitting models,
resulting from a joint fit to all the \chandra\ or \xmm\ data,
 are shown in Fig~\ref{fig_spectra}.
The best-fitting abundance profile, and abundance ratios are given in Fig~\ref{fig_abundance},
while the projected temperature and density profiles are shown in Fig~\ref{fig_kt_rho}.

\section{Mass modelling} \label{sect_mass}
Under the hydrostatic approximation, we transformed the projected density and temperature data into
mass constraints with the entropy-based ``forward fitting'' technique described in 
\citet[][]{humphrey11a,humphrey08a}, which enables tight control over systematic errors,
in comparison with other popular methods, such as ``smoothed inversion'' 
\citep[for a review of mass modelling methods, see][]{buote11a}.
Briefly, the entropy-based 
forward fitting method involves solving the equation of hydrostatic equilibrium to 
compute temperature and density profile models, given parametrized mass and entropy profiles. 
The models were then projected onto the sky and fitted to the projected temperature and density 
profiles, taking into account covariance between the density data points.
As is standard, we assumed spherical symmetry, which does not introduce substantial biases
into the inferred mass distribution \citep{buote11c}.
So as not to violate the Schwarzschild criterion for stability against convection, the 
entropy profile must rise monotonically, so we parametrized it as a constant plus a broken 
powerlaw model. We explored different models for the mass distribution, as discussed below,
and allowed the parameters describing the entropy profile, plus the logarithm of the gas density
at a fiducial radius, and the mass profile parameters to fit freely. Parameter space exploration
employed version 2.7 of the MultiNest Bayesian code\footnote{{http://www.mrao.cam.ac.uk/software/multinest/}} \citep{feroz09a}. Initially we adopted flat priors for each fit parameter, but
subsequently explored the impact of this choice on the fit results (\S~\ref{sect_syserr}).

\begin{figure}
\centering
\includegraphics[width=3.4in]{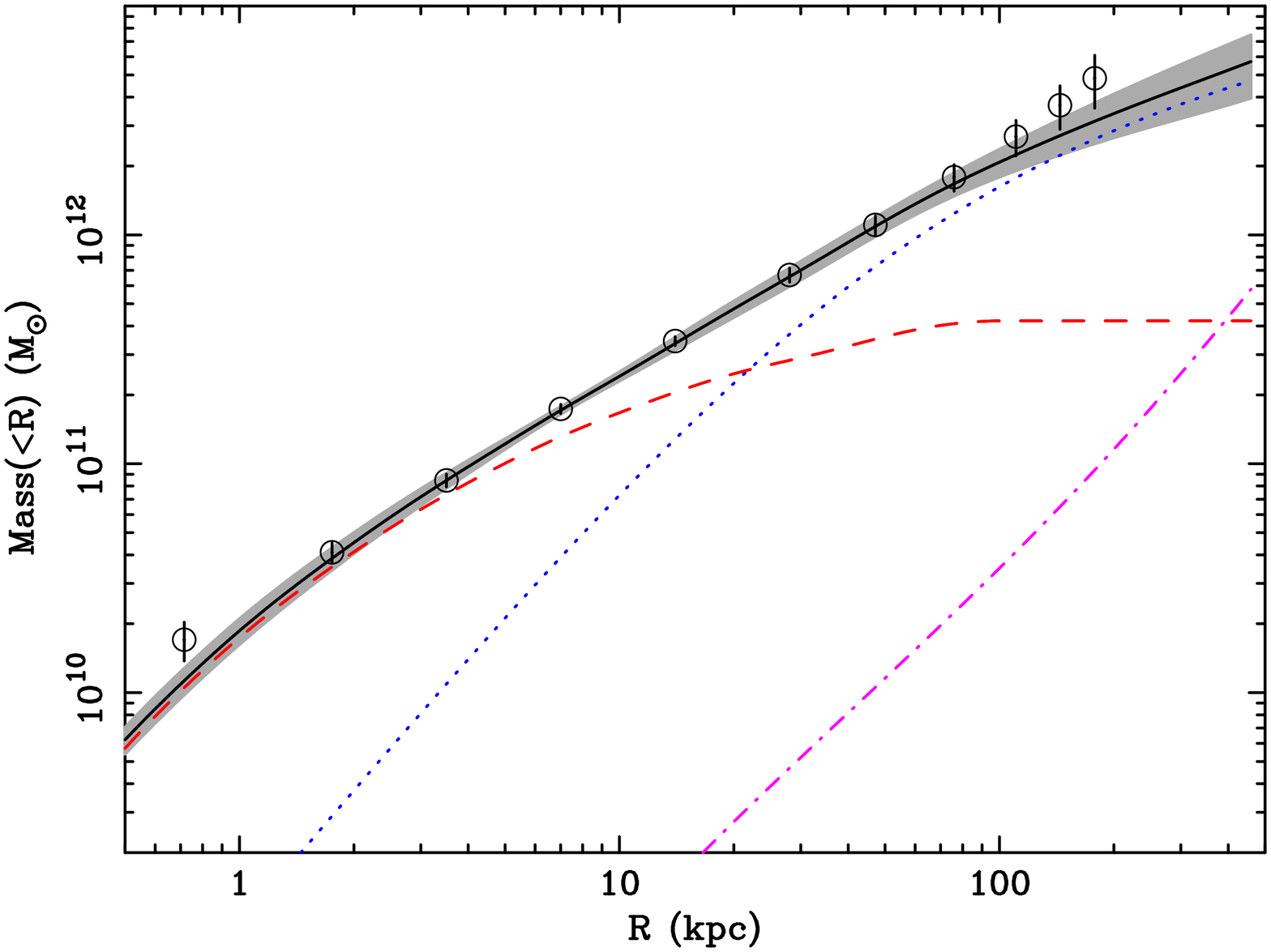}
\caption{Radial mass profile of \src. The solid (black) line
indicates the total enclosed mass (and the grey shaded region
indicates the 1-$\sigma$ error in the total mass distribution),
the dashed (red) line indicates the 
stellar mass, the dotted (blue) line is the dark matter, and the 
dash-dot (magenta) line is the gas mass contribution. 
Overlaid are a set of data-points derived from a more traditional
``smoothed inversion'' approach \citep{buote11a,humphrey09d}.
We stress that the
model is {\em not} fitted to these data, but is derived independently
from the temperature and density data.
\label{fig_mass_profile}}
\end{figure}

For the mass models, we first considered the case of no DM. We included a stellar light component, based on spherically averaging the deprojected, triaxial model for the I-band
light discussed in \S~\ref{sect_optical}. As shown in \citet{buote11c}, the mass inferred
from spherical hydrostatic methods should be very close to the spherically averaged 
true mass.  The mass-to-light (M/L) ratio was allowed to fit freely. We also included a 
(fixed) black hole with mass \mbh=$3\times 10^{8}$\msun, consistent with the \mbh-$\sigma_*$
relation of \citet{gultekin09a}, given the central stellar velocity dispersion
\sigmac$=224$\kms\ \citep{faber89}. 
{The fit to the density and temperature profiles was
poor ($\chi^2$/dof=107/21), as shown in Fig~\ref{fig_kt_rho}. 
Next, we added an NFW \citep{navarro97} DM halo. We allowed $log_{10}$\mvir\ and 
$log_{10}$\cvir\ (the DM halo concentration) to fit freely 
(restricting $11\le log_{10} M_{vir}\le 15$ and $0\le log_{10} c_{vir}\le 2$), 
and obtained a formally 
acceptable fit 
($\chi^2$/dof=29.5/19; Fig~\ref{fig_kt_rho}).
The measured radial mass distribution is 
shown in Fig~\ref{fig_mass_profile}.}

The improvement to the fit when dark matter was 
included was highly significant; the ratio of the Bayesian 
``evidence'' returned for the two cases is $1.8\times 10^{-16}$, implying DM is 
required at 8.2-$\sigma$. In Table~\ref{table_mass}, we 
give the stellar M/L ratio, and show the 
marginalized total mass within $R_\Delta$ for various different 
overdensities ($\Delta$), and the corresponding
concentrations ($c_\Delta=R_\Delta/r_s$;
$r_s$ is the characteristic scale of the NFW model).


In order to explore how \src\ sits on the $\alpha$-\reff\ relation
discovered by \citet{humphrey10a}, we experimented with fitting
the data within the central 50~kpc, under the assumption that the 
mass density profile is given by $\rho \propto r^{-\alpha}$.
The model fits the data comparably to the NFW+stars model discussed
above, confirming that the central part of the mass profile is 
approximately powerlaw in form. The constraints on the mass model
are summarized in Table~\ref{table_mass_slope}. We obtained
$\alpha=1.95^{+0.04}_{-0.06}$, comparable to other systems 
in \citet{humphrey10a} with K-band \reff\ similar
to \src\ (=4.7~kpc, as given in \twomass)\footnote{We used the \twomass\
measurement of \reff\ for \src\ to enable a consistent comparison
with the measurements from \citet{humphrey10a}.}.
\renewcommand{\tabcolsep}{0.1mm}
\begin{deluxetable*}{llllllll}
\tablecaption{Mass results and error budget\label{table_mass}\vspace{-10pt}}
\tabletypesize{\scriptsize}
\centering
\tablehead{
\colhead{Test} & \colhead{M$_*$/\li} & \colhead{log \mtwentyfive} & \colhead{log $c_{2500}$}& \colhead{log \mfive} & \colhead{log $c_{500}$} & \colhead{log \mvir} & \colhead{log \cvir} \\ 
& \colhead{\msun\lsun$^{-1}$}& \colhead{[\msun]} & & \colhead{[\msun]} & & \colhead{[\msun]}  
}
\startdata
Marginalized & $2.49^{+0.26}_{-0.42}$& $12.40 \pm 0.08$& $0.61^{+0.16}_{-0.20}$& $12.59 \pm 0.11$& $0.90^{+0.15}_{-0.18}$& $12.73^{+0.15}_{-0.10}$& $1.19^{+0.14}_{-0.17}$\\
 Best-fit& $(2.55)$& $(12.40)$& $(0.58)$& $(12.59)$& $(0.88)$& $(12.76)$& $(1.16)$\\\hline
$\Delta$DM profile&  $+0.34$  $\left( \pm {0.37}\right)$  &  $+0.02$  $\left( \pm {0.06}\right)$  & \ldots  &  $+0.16$  $\left( \pm {0.06}\right)$  & \ldots &  $+0.36$  $\left( \pm {0.07}\right)$  &  \ldots\\
$\Delta$AC&  $-0.652$  $\left( \pm {0.37}\right)$  &  $+0.008$  $\left( \pm {0.09}\right)$  &  $-0.086$  $\left( ^{+0.21}_{-0.28}\right)$  &  $+0.003$  $\left( ^{+0.15}_{-0.11}\right)$  &  $-0.073$  $\left( ^{+0.19}_{-0.27}\right)$  &  $+0.02$  $\left( ^{+0.20}_{-0.12}\right)$  &  $-0.072$  $\left( ^{+0.18}_{-0.26}\right)$ \\
$\Delta$Stars &  \ldots  &  $-0.045$  $\left( \pm {0.08}\right)$  &  $+0.18$  $\left( ^{+0.14}_{-0.23}\right)$  &  $-0.043$  $\left( \pm {0.11}\right)$  &  $+0.18$  $\left( ^{+0.13}_{-0.21}\right)$  &  $\pm 0.03$  $\left( \pm {0.13}\right)$  &  $+0.18$  $\left( ^{+0.12}_{-0.21}\right)$ \\
$\Delta$H.E.&  $+0.83$  $\left( ^{+0.37}_{-0.53}\right)$  &  $+0.16$  $\left( \pm {0.09}\right)$  &  $+0.02$  $\left( \pm {0.20}\right)$  &  $+0.15$  $\left( \pm {0.12}\right)$  &  $^{+0.00}_{-0.03}$  $\left( ^{+0.21}_{-0.15}\right)$  &  $+0.17$  $\left( \pm {0.13}\right)$  &  $^{+0.00}_{-0.04}$  $\left( ^{+0.21}_{-0.14}\right)$ \\
$\Delta$Background&  $\pm 0.13$  $\left( ^{+0.35}_{-0.50}\right)$  &  $^{+0.16}_{-0.05}$  $\left( \pm {0.10}\right)$  &  $^{+0.08}_{-0.12}$  $\left( ^{+0.21}_{-0.29}\right)$  &  $^{+0.20}_{-0.07}$  $\left( \pm {0.14}\right)$  &  $^{+0.04}_{-0.11}$  $\left( \pm {0.21}\right)$  &  $^{+0.23}_{-0.03}$  $\left( \pm {0.16}\right)$  &  $^{+0.04}_{-0.11}$  $\left( \pm {0.20}\right)$ \\
$\Delta$SWCX&  $-0.095$  $\left( \pm {0.42}\right)$  &  $-0.090$  $\left( ^{+0.12}_{-0.07}\right)$  &  $+0.11$  $\left( ^{+0.17}_{-0.22}\right)$  &  $-0.122$  $\left( ^{+0.16}_{-0.09}\right)$  &  $+0.10$  $\left( ^{+0.16}_{-0.21}\right)$  &  $-0.104$  $\left( ^{+0.17}_{-0.10}\right)$  &  $+0.10$  $\left( ^{+0.16}_{-0.21}\right)$ \\
$\Delta$Instrument&  $+0.22$  $\left( \pm {0.31}\right)$  &  $+0.08$  $\left( ^{+0.16}_{-0.11}\right)$  &  $-0.243$  $\left( \pm {0.28}\right)$  &  $+0.15$  $\left( \pm {0.22}\right)$  &  $-0.219$  $\left( \pm {0.25}\right)$  &  $+0.18$  $\left( ^{+0.30}_{-0.21}\right)$  &  $-0.206$  $\left( ^{+0.21}_{-0.26}\right)$ \\
$\Delta$Fit radius&  $+0.06$  $\left( \pm {0.27}\right)$  &  $+0.20$  $\left( ^{+0.10}_{-0.13}\right)$  &  $-0.232$  $\left( \pm {0.20}\right)$  &  $+0.24$  $\left( \pm {0.17}\right)$  &  $-0.212$  $\left( \pm {0.18}\right)$  &  $+0.28$  $\left( \pm {0.19}\right)$  &  $-0.198$  $\left( \pm {0.17}\right)$ \\
$\Delta$3d &  $-0.353$  $\left( \pm {0.53}\right)$  &  $-0.079$  $\left( ^{+0.19}_{-0.10}\right)$  &  $+0.03$  $\left( ^{+0.33}_{-0.20}\right)$  &  $-0.096$  $\left( ^{+0.24}_{-0.13}\right)$  &  $+0.01$  $\left( ^{+0.33}_{-0.17}\right)$  &  $-0.084$  $\left( ^{+0.26}_{-0.09}\right)$  &  $+0.03$  $\left( ^{+0.31}_{-0.19}\right)$ \\
$\Delta$Fit priors&  $+0.39$  $\left( \pm {0.21}\right)$  &  $+0.06$  $\left( \pm {0.08}\right)$  &  $-0.296$  $\left( \pm {0.14}\right)$  &  $+0.15$  $\left( \pm {0.13}\right)$  &  $-0.275$  $\left( \pm {0.13}\right)$  &  $+0.19$  $\left( \pm {0.14}\right)$  &  $-0.261$  $\left( \pm {0.13}\right)$ \\
$\Delta$Spectral&  $-0.194$  $\left( ^{+0.24}_{-0.33}\right)$  &  $-0.043$  $\left( \pm {0.06}\right)$  &  $+0.04$  $\left( \pm {0.15}\right)$  &  $-0.051$  $\left( \pm {0.09}\right)$  &  $+0.03$  $\left( \pm {0.14}\right)$  &  $-0.027$  $\left( \pm {0.10}\right)$  &  $+0.03$  $\left( \pm {0.13}\right)$ \\
$\Delta$Entropy &  $+0.03$  $\left( ^{+0.23}_{-0.45}\right)$  &  $-0.006$  $\left( ^{+0.10}_{-0.07}\right)$  &  $-0.010$  $\left( \pm {0.19}\right)$  &  $-0.018$  $\left( ^{+0.15}_{-0.09}\right)$  &  $-0.010$  $\left( \pm {0.17}\right)$  &  $+0.01$  $\left( ^{+0.16}_{-0.12}\right)$  &  $-0.028$  $\left( \pm {0.17}\right)$ \\
$\Delta$Weighting&  $-0.158$  $\left( \pm {0.44}\right)$  &  $\pm 0$  $\left( \pm {0.09}\right)$  &  $+0.04$  $\left( ^{+0.19}_{-0.23}\right)$  &  $-0.025$  $\left( ^{+0.16}_{-0.10}\right)$  &  $+0.005$  $\left( \pm {0.19}\right)$  &  $-0.015$  $\left( ^{+0.19}_{-0.11}\right)$  &  $+0.03$  $\left( \pm {0.19}\right)$ \\
$\Delta$Distance &  $^{+0.59}_{-0.45}$  $\left( ^{+0.36}_{-0.49}\right)$  &  $-0.054$  $\left( ^{+0.11}_{-0.07}\right)$  &  $\pm 0.08$  $\left( ^{+0.20}_{-0.24}\right)$  &  $-0.080$  $\left( ^{+0.16}_{-0.08}\right)$  &  $\pm 0.08$  $\left( ^{+0.18}_{-0.22}\right)$  &  $^{+0.00}_{-0.05}$  $\left( ^{+0.19}_{-0.11}\right)$  &  $\pm 0.08$  $\left( ^{+0.17}_{-0.21}\right)$ \\
$\Delta$Covariance &  $+0.16$  $\left( ^{+0.35}_{-0.82}\right)$  &  $^{+0.05}_{-0.02}$  $\left( ^{+0.13}_{-0.09}\right)$  &  $^{+0.07}_{-0.18}$  $\left( ^{+0.23}_{-0.32}\right)$  &  $^{+0.09}_{-0.05}$  $\left( ^{+0.20}_{-0.12}\right)$  &  $^{+0.07}_{-0.16}$  $\left( ^{+0.22}_{-0.31}\right)$  &  $^{+0.13}_{-0.04}$  $\left( ^{+0.24}_{-0.12}\right)$  &  $^{+0.06}_{-0.15}$  $\left( ^{+0.21}_{-0.30}\right)$ 
\enddata
\tablecomments{{Marginalized values and 1-$\sigma$ confidence regions for the stellar 
mass-to-light (M$_*$/\lk) ratio and the enclosed mass and concentration measured at various overdensities. 
Since the best-fitting parameters need not be identical to the marginalized
values, we also list the best-fitting values for each parameter (in parentheses). }
In addition to the statistical errors, we also show estimates of the error budget
from possible sources of systematic uncertainty.
We consider a range of different
systematic effects, which are described in detail in \S~\ref{sect_syserr}; 
specifically we evaluate the effect of the
choice of dark matter halo model ($\Delta$DM), adiabatic contraction ($\Delta$AC),
treatment of the stellar light ($\Delta$Stars), plausible deviations from hydrostatic
equilibrium ($\Delta$H.E.), treatment of the background ($\Delta$Background) and 
the Solar wind charge exchange X-ray component ($\Delta$SWCX), the instrumental
inter-calibration ($\Delta$Instrument), the radial coverage of the data being 
fitted ($\Delta$Fit radius), 
deprojection ($\Delta$3d), priors on the model parameters 
($\Delta$Fit priors), spectral fitting choices ($\Delta$Spectral),
the parameterization of the entropy model ($\Delta$Entropy),
removing the emissivity correction ($\Delta$Weighting), 
distance uncertainties ($\Delta$Distance), and the treatment of covariance between the 
temperature and density data-points ($\Delta$Covariance).  
We list the change
in the marginalized value of each parameter for every test and, in parentheses,
the statistical uncertainty on the parameter determined from the test.
Note that the systematic error estimates should {\em not}
in general be added in quadrature with the statistical error. Since there 
is no theoretical interest in the distribution of scale radii for the cored logarithmic
DM model, we do not include changes on the concentration in our error budget for this 
choice of profile ($\Delta$DM profile). Likewise, the tests involving different stellar
light modelling approaches ($\Delta$Stars) involved using light profiles 
from different optical filter, and so we omit the M$_*$/L$_I$ ratio from the error
budget for that choice.}
\end{deluxetable*}
\renewcommand{\tabcolsep}{0.3mm}
\begin{deluxetable}{lll}
\tablecaption{Mass slope results and error budget\label{table_mass_slope}}
\tabletypesize{\scriptsize}
\centering
\tablehead{
\colhead{Test} & \colhead{log M$_{75}$} & \colhead{$\alpha$} \\
& \colhead{[\msun]} & }
\startdata
$\Delta$H.E.&  $+0.03$  $\left( \pm {0.05}\right)$  &  $+0.04$  $\left( \pm {0.05}\right)$ \\
$\Delta$Background&  $^{+0.02}_{-0.03}$  $\left( \pm {0.06}\right)$  &  $^{+0.05}_{-0.02}$  $\left( \pm {0.05}\right)$ \\
$\Delta$SWCX&  $+0.01$  $\left( ^{+0.06}_{-0.05}\right)$  &  $-0.010$  $\left( \pm {0.05}\right)$ \\
$\Delta$Instrument&  $^{+0.12}_{-0.03}$  $\left( \pm {0.06}\right)$  &  $^{+0.03}_{-0.10}$  $\left( \pm {0.06}\right)$ \\
$\Delta$Fit radius&  $+0.03$  $\left( ^{+0.04}_{-0.03}\right)$  &  $-0.028$  $\left( ^{+0.03}_{-0.05}\right)$ \\
$\Delta$3d &  $+0.05$  $\left( \pm {0.07}\right)$  &  $-0.075$  $\left( \pm {0.06}\right)$ \\
$\Delta$Fit priors&  $^{+0.01}_{-0.00}$  $\left( \pm {0.06}\right)$  &  $^{+0.01}_{-0.01}$  $\left( \pm {0.06}\right)$ \\
$\Delta$Spectral&  $-0.020$  $\left( \pm {0.04}\right)$  &  $+0.008$  $\left( \pm {0.04}\right)$ \\
$\Delta$Entropy &  $-0.023$  $\left( \pm {0.05}\right)$  &  $+0.02$  $\left( \pm {0.05}\right)$ \\
$\Delta$Weighting&  $+0.03$  $\left( ^{+0.05}_{-0.06}\right)$  &  $-0.033$  $\left( ^{+0.07}_{-0.04}\right)$ \\
$\Delta$Distance &  $-0.037$  $\left( ^{+0.05}_{-0.06}\right)$  &  $+0.03$  $\left( \pm {0.05}\right)$ \\
$\Delta$Covariance &  $^{+0.05}_{-0.03}$  $\left( \pm {0.07}\right)$  &  $\pm 0.04$  $\left( ^{+0.06}_{-0.10}\right)$ \\
\enddata
\tablecomments{Marginalized values and 1-$\sigma$ confidence regions for the 
enclosed mass at 75~kpc ($M_{75}$) and the negative logarithmic
slope of the mass profile ($\alpha$), when fitting only a single 
powerlaw to the total mass distribution. 
We also provide the best-fitting parameters in parentheses, and a breakdown 
of possible sources of systematic uncertainty, following Table~\ref{table_mass}.}
\end{deluxetable}

\begin{deluxetable*}{llllrrr}
\tablecaption{Baryon fraction results and error budget\label{table_fb}}
\tabletypesize{\scriptsize}
\centering
\tablehead{
\colhead{Test} & \colhead{$f_{g,2500}$} & \colhead{$f_{g,500}$} & \colhead{$f_{g,vir}$}& \colhead{$f_{b,2500}$} & \colhead{$f_{b,500}$} & \colhead{$f_{b,vir}$}\\
}
\startdata
Marginalized & $0.019 \pm 0.002$& $0.040^{+0.006}_{-0.008}$& $0.10 \pm 0.02$& $0.16 \pm 0.03$& $0.13 \pm 0.03$& $0.16 \pm 0.03$\\
Best-fit& $(0.018)$& $(0.039)$& $(0.095)$& $(0.178)$& $(0.142)$& $(0.165)$\\\hline
$\Delta$DM profile&  $\pm 0$  $\left( \pm {0.002}\right)$  &  $-0.007$  $\left( \pm {0.005}\right)$  &  $-0.029$  $\left( \pm {0.01}\right)$  &  $+0.005$  $\left( \pm {0.04}\right)$  &  $-0.021$  $\left( \pm {0.02}\right)$  &  $-0.056$  $\left( ^{+0.02}_{-0.02}\right)$ \\
$\Delta$AC&  $\pm 0$  $\left( \pm {0.002}\right)$  &  $-0.001$  $\left( ^{+0.01}_{-0.01}\right)$  &  $-0.007$  $\left( \pm {0.02}\right)$  &  $-0.042$  $\left( \pm {0.02}\right)$  &  $-0.026$  $\left( ^{+0.02}_{-0.02}\right)$  &  $-0.023$  $\left( \pm {0.03}\right)$ \\
$\Delta$Stars &  $\pm 0$  $\left( \pm {0.003}\right)$  &  $\pm 0$  $\left( \pm {0.01}\right)$  &  $+0.009$  $\left( \pm {0.02}\right)$  &  $-0.085$  $\left( ^{+0.03}_{-0.02}\right)$  &  $-0.049$  $\left( \pm {0.02}\right)$  &  $-0.029$  $\left( \pm {0.03}\right)$ \\
$\Delta$H.E.&  $-0.003$  $\left( \pm {0.002}\right)$  &  $-0.007$  $\left( \pm {0.01}\right)$  &  $-0.015$  $\left( \pm {0.02}\right)$  &  $\pm 0.02$  $\left( \pm {0.03}\right)$  &  $^{+0.01}_{-0.01}$  $\left( \pm {0.03}\right)$  &  $^{+0.01}_{-0.02}$  $\left( \pm {0.04}\right)$ \\
$\Delta$Background&  $-0.007$  $\left( \pm {0.003}\right)$  &  $^{+0.002}_{-0.018}$  $\left( \pm {0.01}\right)$  &  $^{+0.01}_{-0.05}$  $\left( \pm {0.03}\right)$  &  $^{+0.01}_{-0.04}$  $\left( \pm {0.04}\right)$  &  $^{+0.02}_{-0.04}$  $\left( \pm {0.03}\right)$  &  $^{+0.02}_{-0.06}$  $\left( \pm {0.05}\right)$ \\
$\Delta$SWCX&  $+0.001$  $\left( \pm {0.003}\right)$  &  $+0.004$  $\left( \pm {0.01}\right)$  &  $+0.007$  $\left( \pm {0.03}\right)$  &  $+0.006$  $\left( ^{+0.04}_{-0.04}\right)$  &  $+0.009$  $\left( ^{+0.04}_{-0.03}\right)$  &  $+0.02$  $\left( \pm {0.05}\right)$ \\
$\Delta$Instrument&  $\pm 0$  $\left( ^{+0.002}_{-0.004}\right)$  &  $-0.006$  $\left( ^{+0.01}_{-0.01}\right)$  &  $-0.023$  $\left( ^{+0.03}_{-0.03}\right)$  &  $-0.029$  $\left( \pm {0.03}\right)$  &  $-0.025$  $\left( ^{+0.03}_{-0.04}\right)$  &  $-0.046$  $\left( \pm {0.05}\right)$ \\
$\Delta$Fit radius&  $-0.004$  $\left( ^{+0.003}_{-0.003}\right)$  &  $-0.011$  $\left( ^{+0.01}_{-0.01}\right)$  &  $-0.059$  $\left( ^{+0.04}_{-0.02}\right)$  &  $-0.056$  $\left( ^{+0.03}_{-0.02}\right)$  &  $-0.049$  $\left( \pm {0.03}\right)$  &  $-0.071$  $\left( \pm {0.04}\right)$ \\
$\Delta$3d &  $+0.004$  $\left( \pm {0.006}\right)$  &  $+0.02$  $\left( ^{+0.02}_{-0.03}\right)$  &  $+0.05$  $\left( ^{+0.06}_{-0.11}\right)$  &  $-0.036$  $\left( ^{+0.05}_{-0.03}\right)$  &  $+0.002$  $\left( \pm {0.05}\right)$  &  $+0.04$  $\left( ^{+0.08}_{-0.12}\right)$ \\
$\Delta$Fit priors&  $\pm 0$  $\left( \pm {0.003}\right)$  &  $-0.005$  $\left( \pm {0.01}\right)$  &  $-0.019$  $\left( \pm {0.02}\right)$  &  $^{+0.00}_{-0.01}$  $\left( \pm {0.03}\right)$  &  $^{+0.00}_{-0.01}$  $\left( \pm {0.03}\right)$  &  $-0.030$  $\left( \pm {0.04}\right)$ \\
$\Delta$Spectral&  $+0.002$  $\left( \pm {0.002}\right)$  &  $+0.004$  $\left( ^{+0.005}_{-0.008}\right)$  &  $+0.01$  $\left( ^{+0.02}_{-0.02}\right)$  &  $\pm 0$  $\left( \pm {0.02}\right)$  &  $+0.01$  $\left( ^{+0.02}_{-0.02}\right)$  &  $+0.02$  $\left( ^{+0.03}_{-0.03}\right)$ \\
$\Delta$Entropy &  $\pm 0$  $\left( ^{+0.003}_{-0.002}\right)$  &  $\pm 0$  $\left( \pm {0.01}\right)$  &  $-0.003$  $\left( ^{+0.02}_{-0.03}\right)$  &  $-0.013$  $\left( ^{+0.03}_{-0.02}\right)$  &  $-0.005$  $\left( \pm {0.03}\right)$  &  $-0.009$  $\left( \pm {0.03}\right)$ \\
$\Delta$Weighting&  $-0.003$  $\left( ^{+0.003}_{-0.002}\right)$  &  $-0.006$  $\left( \pm {0.01}\right)$  &  $-0.012$  $\left( ^{+0.02}_{-0.02}\right)$  &  $-0.019$  $\left( \pm {0.03}\right)$  &  $-0.012$  $\left( \pm {0.03}\right)$  &  $-0.015$  $\left( ^{+0.03}_{-0.04}\right)$ \\
$\Delta$Distance &  $\pm 0.003$  $\left( ^{+0.003}_{-0.002}\right)$  &  $^{+0.004}_{-0.006}$  $\left( ^{+0.01}_{-0.01}\right)$  &  $\pm 0.01$  $\left( \pm {0.02}\right)$  &  $\pm 0.03$  $\left( \pm {0.03}\right)$  &  $^{+0.02}_{-0.02}$  $\left( \pm {0.03}\right)$  &  $^{+0.03}_{-0.01}$  $\left( ^{+0.04}_{-0.04}\right)$ \\
$\Delta$Covariance &  $\pm 0$  $\left( \pm {0.002}\right)$  &  $^{+0.002}_{-0.004}$  $\left( ^{+0.01}_{-0.01}\right)$  &  $^{+0.005}_{-0.012}$  $\left( ^{+0.03}_{-0.03}\right)$  &  $-0.017$  $\left( \pm {0.03}\right)$  &  $-0.010$  $\left( \pm {0.03}\right)$  &  $-0.015$  $\left( \pm {0.04}\right)$ \\
\enddata
\tablecomments{Marginalized values and 1-$\sigma$ confidence regions for the gas 
fraction ($f_{g,\Delta}$) and baryon fraction ($f_{b,\Delta}$) measured at various overdensities
($\Delta$). We also provide the best-fitting parameters in parentheses, and a breakdown 
of possible sources of systematic uncertainty, following Table~\ref{table_mass}.}
\end{deluxetable*}


{
\section{Systematic Error Budget} \label{sect_syserr} 
In this section, we address the sensitivity of our results to
various data analysis choices that were made.
Since it is generally impractical to express these assumptions
through an additional model parameter over which one can
marginalize, we adopted the pragmatic approach of exploring how
our results changed if the assumptions were adjusted in an
arbitrary, but representative, way. We focused on those 
systematic effects likely to have the greatest impact on our 
conclusions. In Tables~\ref{table_mass}--\ref{table_fb}, 
we list the change in the marginalized value of each parameter.
We discuss each test in more detail below. In summary, most of the inferred
systematic errors are comparable to the statistical errors. 
For a couple of the tests, marginally significant ($\sim$2-$\sigma$) increases 
in \mvir, and corresponding reductions in \fbvir\ are seen (by as much as $\sim$0.07),
but our conclusions are not strongly affected.

\subsection{Dark matter halo} \label{sect_syserr_mass} 
The accurate computation of the gas distribution out to \rvir\ is 
contingent upon the accurate modelling of the gravitating mass distribution.
While the NFW DM halo model is well-motivated theoretically, we also
considered the ``cored logarithmic'' model that is sometimes used
in stellar dynamical studies \citep[\eg][]{binney08a,shen10a}.
This model requires a slightly
higher \mvir, and correspondingly lower \fbvir\ (``$\Delta$DM profile'' in 
Tables~\ref{table_mass} and \ref{table_fb}). Although we cannot distinguish
between the NFW and cored logarithmic model on the basis of $\chi^2$ alone,
the ratio of the Bayesian evidence ($7\times 10^{-4}$) implies that the 
cored logarithmic model, with the adopted priors 
(a flat prior on the 
asymptotic circular velocity, between 10 and 2000~$km\ s^{-1}$, and a flat 
prior on $log_{10} r_c$, where $r_c$ is the core radius, over the range
$0\le log_{10} r_c \le 3$.) is a poorer description of the data at 
$\sim$3.4-$\sigma$.

A theoretical modification to the DM profile is expected to arise
from adiabatic contraction \citep{blumenthal86a,gnedin07a,abadi09a}, 
which causes the halo to become cuspier 
due to the gravitational influence of the baryons. Modifying the 
NFW profile with the algorithm of \citet{gnedin04a}\footnote{Using the CONTRA code publicly available from http://www.astro.lsa.umich.edu/$\sim$ognedin/contra/}, 
has only a very slight effect on the best-fitting mass model 
(``$\Delta$AC'' in Tables~\ref{table_mass} and \ref{table_fb}), except for the
stellar M/L ratio, which is reduced to make room for the increased  DM 
fraction predicted in the inner parts of the galaxy.

\subsection{Stellar light} \label{sect_syserr_stars} 
Accurately decomposing the gravitating mass distribution into the 
luminous and dark components requires an accurate model for the (deprojected)
stellar light. In the case of \src, we
used a triaxial deprojection procedure (\S~\ref{sect_optical}).
Although we averaged this profile
spherically for use with our X-ray modelling code, we have 
previously shown that 
this approach does not generally introduce significant biases 
\citep{buote11b,buote11c}. Nevertheless, the deprojection procedure may not
be unique, and so it is important to explore how sensitive our results 
are to the details of the stellar light modelling. 

We experimented with three alternative prescriptions for deprojecting the 
stellar light. First, we fitted the B-band surface brightness profiles
along the major and minor axis published by 
\citet{capaccioli88a} with a model comprising 6 multiple, concentric 
3-dimensional Gaussian density distributions, that were projected
onto the sky, assuming an edge-on, oblate geometry. This ignores
the isophotal twist. Second, we fitted the K-band \twomass\ image
in the central $\sim$4\arcmin\ region with a model comprising an
elliptical Sersic model, which we deprojected with the 
formula of \citet{prugniel97a}, assuming an edge-on oblate spheroidal
geometry. Finally, we adopted a spherical, deprojected de Vaucouleurs
model, the effective radius and luminosity of which were set to 
match the catalogued \twomass\ K-band values. We found that the
treatment of the stellar light primarily only affected the 
total stellar mass, so that \fbtwentyfive\ is very sensitive to
this choice. However, by \rvir\ the impact is much less
significant  (``$\Delta$Stars'' in Tables~\ref{table_mass}
and \ref{table_fb}).

\subsection{Hydrostatic equilibrium} \label{sect_syserr_he}
Although the gas in morphologically relaxed
early-type galaxies is expected to be close to hydrostatic,
recent work suggests that 
 a small amount (\ltsim 30\%) of nonthermal support is present
in the very central parts of some galaxies 
\citep[\eg][]{churazov08a,das10a,humphrey12b}. To investigate
whether deviations from hydrostatic equilibrium at this 
level would quantitatively affect our conclusions, we modified
the hydrostatic equation used in our modelling code to include
a plausible nonthermal component.
{First we considered a nonthermal pressure fraction profile similar to that
inferred for the galaxy NGC\thin 4649, which was  
fixed at $\sim$25\%\ at the centre and fell to $\sim$10\%\ by 20~kpc,
vanishing outside 30~kpc \citep{humphrey12b}. This is similar 
to the implied nonthermal pressure profiles that have been 
inferred in (albeit a handful of) other systems \citep[\eg][]{das10a}.
Since the nonthermal pressure is most important in the central
few kpc, we found that adding this component did not significantly
affect the global parameters of the system, although the stellar
M/L ratio was increased by $\sim$18\%\ to compensate.
As a more extreme alternative, we also considered a uniform nonthermal
pressure fraction fixed at 25\%, similar in magnitude to that inferred
from previous studies of the central parts of of galaxies 
\citep[\eg][]{churazov08a,churazov10a,humphrey09d,humphrey12b,das10a}. 
In this case, the stellar M/L ratio was increased more substantially
(by $\sim$33\%), and the global mass raised by $\sim$0.15~dex. 
Conversely, the baryon and gas fractions at fixed overdensity were
found to be relatively insensitive to this choice, since {\em both}
stellar and gas mass are increased, while the slightly larger \rfive,
for example, encloses more gas
(``$\Delta$H.E.'' in Tables~\ref{table_mass}--\ref{table_fb}).}

\subsection{Background} \label{sect_syserr_background}
Since the data were background-dominated in the outer \xmm\ 
annuli, the treatment of the background was a potentially serious
source of systematic uncertainty. To investigate the extent to
which our results are sensitive to this, we explored a number of 
different choices in our background treatment.
First, for the \chandra\ data, we adopted the 
standard blank-field events files distributed with the CALDB to
extract a background spectrum for each annulus.
Since the blank-field  files for each CCD have different 
exposures, spectra were accumulated for each CCD individually, scaled to a
common exposure time and then added. The spectra were renormalized 
to match the observed count-rate in the 9--12~keV band. These ``template'' 
spectra were then used as a background in \xspec, and the background model components were
omitted from our fit. This did not strongly affect our conclusions.

Since the non X-ray component dominates the \xmm\ background for much
of the band-pass, we explored an  alternative means of accounting for
it. Specifically, we adopted the ESAS algorithm
\citep{kuntz08a,snowden08a}. This involves choosing non X-ray 
background template files that match the count-rate and hardness ratios of 
photons in the unexposed portions of the CCDs. We added the 
X-ray background and instrumental lines, as in our standard 
modelling procedure. Using this approach, we found a modest
reduction in the gas density at all radii (Fig~\ref{fig_kt_rho}). In the 
innermost regions this arises due to a slightly higher best-fitting 
abundance (\zfe$\simeq 0.6$). Although this lowers the baryon fraction
slightly (by 4\%\ at \rfive), this change is comparable to the statistical 
error. Next, we explored the impact of varying the X-ray background. 
We varied the slope of the cosmic X-ray component (powerlaw model) by 
$\pm5$\%, changed kT of the local hot bubble component to
0.1~keV, and varied kT for the galactic hot gas component between
0.18~keV and 0.24~keV, which span the range of temperatures obtained
from fitting nearby blank-sky fields (\S~\ref{sect_xmm}).
Although these choices had a measurable impact on the global parameters
we derived for \src, our conclusions were largely unaffected. 
We summarize the results in Tables~\ref{table_mass}--\ref{table_fb} 
(``$\Delta$Background'').

\subsubsection{Solar Wind Charge Exchange} \label{sect_syserr_swcx}
A potentially relevant, time variable, soft background component can
arise due to the interaction of the Solar wind with interstellar material
and the Earth's exosphere. To explore the possible importance of this
``Solar Wind Charge Exchange'' (SWCX) component, we used data from
the WIND-SWE experiment \citep{ogilvie95a}\footnote{\href{http://web.mit.edu/space/www/wind\_data.html\#Protons}{http://web.mit.edu/space/www/wind\_data.html\#Protons}}
to identify periods of strong Solar wind activity. 
Following \citet{snowden04a}, we assumed the SWCX component
is negligible for a Solar wind proton flux level measured to be
\ltsim $3\times 10^8 cm^{-2}\ s^{-1}$, and excised all 
other data. While this eliminated $\sim$30\%\ of both the \chandra\
and \xmm\ exposures, we found that this choice only had a minimal impact on the 
derived parameters (``$\Delta$SWCX'' in Tables~\ref{table_mass}--\ref{table_fb}),
indicating that the SWCX is not a significant problem in our analysis. 

\subsection{Instrumental inter-calibration} \label{sect_syserr_instrument}
In our default \xmm\ analysis, we considered only the MOS1 and MOS2
instruments, as we have found the calibration of the PN to be more
uncertain in the low-temperature (kT$\sim$0.5~keV) regime. Nevertheless,
given its large collecting area, it is important to investigate whether
the PN data can add information in the low surface brightness outer regions
of the galaxy. To explore this, we first fitted the PN spectra, similarly to
the MOS data, to obtain the gas  temperature and density profile out to 
$\sim$200~kpc. We then fitted these profiles in tandem with the \chandra\ data.
Although the derived \mvir\ was slightly larger 
(and, consequently, \fgas\ and \fb\ values were smaller) than our best fitting
case, the results were mostly consistent within errors
(``$\Delta$Instrument'' in Tables~\ref{table_mass}--\ref{table_fb}).

\subsection{Radial coverage} \label{sect_syserr_rmax}
The constraints on the global galaxy properties, especially
those derived from a full mass decomposition, can be sensitive
to the radial range being fitted \citep[\eg][]{gastaldello07a}.
To explore the sensitivity of our results to this choice, we restricted
the radial range in our fit to \ltsim 100~kpc (by excluding the outer
three \xmm\ datapoints). This increased the mass of the system by 
0.28~dex (simultaneously reducing \fgas\ and \fb), but only at the expense 
of larger error bars (``$\Delta$Fit radius'' in Tables~\ref{table_mass} and \ref{table_fb}). 

For the powerlaw fits to the total mass distribution, 
by default we excluded all data outside $\sim$50~kpc. Since the 
mass profile must deviate from a pure powerlaw at large scales, we explored
the impact on our results of extending to slightly larger scales in the fit 
(out to $\sim$70~kpc). This did not affect our results significantly
(``$\Delta$Fit radius'' in Table~\ref{table_mass_slope}).


\subsection{Projection/ Deprojection} \label{sect_syserr_3d}
In this work, we fitted the projected, rather than the deprojected
data. To determine the mass profile, we 
modelled the projected temperature and density in 
each annulus by evaluating the hydrostatic model for the temperature
and density in three dimensions, and projecting it onto the line of sight.
In general, this procedure leads to smaller statistical error bars,
but may introduce additional systematic uncertainties related to how
the projected quantities are computed \citep[\eg][]{gastaldello07a}.
To explore the likely impact of these effects, we performed
a spherical deprojection by adding multiple ``vapec'' plasma models
in each annulus, with the relative normalizations tied 
appropriately \citep[\eg][]{kriss83a}. While functionally 
identical to the ``projct'' model in \xspec, this allowed data from both MOS
detectors to be fitted simultaneously. Given the amplification of 
noise by the deprojection procedure (\eg\ \S~3.3 of \citealt{buote00c};
\citealt{finoguenov99}), we excluded the (noisy) outer three \xmm\ annuli
from this analysis. To account for emission projected into the line of 
sight from regions outside the outermost annuli,
we added an APEC plasma component to the spectral model
with an abundance 0.2 (consistent with the outermost annuli) 
and the temperature and normalization determined in each annulus
from projecting onto the line of sight the best-fitting gas temperature and 
density models evaluated
beyond the outer bin. 

Although we found no evidence of biases when comparing the projected and 
deprojected results, the reduction in radial range fitted, in conjunction with the
amplification of error bars during deprojection, resulted in much poorer constraints
on the global quantities computed within \rvir. Nevertheless, at smaller
scales (\eg\ \rtwentyfive), the results were much more robust (``$\Delta$3d'' in 
Tables~\ref{table_mass}--\ref{table_fb}).


\subsection{Priors} \label{sect_syserr_priors}
Since the choice of priors on the various parameters is arbitrary in our analysis,
it is important to determine to what extent they could affect our conclusions.
To do this, we replaced each arbitrary choice in turn with an alternative, reasonable prior.
Specifically, for each parameter describing the entropy profile, we switched from a 
flat prior on that parameter to a flat prior on its logarithm. 
We used a flat prior on the DM halo mass, rather than on its logarithm,
 and, instead of the flat prior on $\log c_{DM}$, we adopted the distribution of 
c around M found by \citet{buote07a} as a (Gaussian) prior.
We also replaced the flat prior on the 
$M_*/L_I$ ratio with a Gaussian prior, corresponding to the 
best-fitting M/L ratio and error-bar derived from fitting the 
published Lick indices (\S~\ref{sect_he}).
For the single powerlaw mass model fits (Table~\ref{table_mass_slope}), we 
used a flat prior on $M_{75}$, rather than its logarithm, and a flat prior
on the logarithm of $\alpha$. 
The effect of these choices is comparable to the statistical errors on each derived
parameter (``$\Delta$Fit priors'' in Tables~\ref{table_mass} and \ref{table_fb}).

\subsection{Other tests}
We here outline the remaining tests we carried out, as summarized
in Tables~\ref{table_mass} and \ref{table_fb}. First of all, to 
assess the sensitivity of our results to the choice of plasma code employed,
we experimented by replacing the APEC plasma model with a MEKAL model.
This had very little impact on our conclusions (``$\Delta$Spectral'' 
in Tables~\ref{table_mass}--\ref{table_fb}).

In order to explore the sensitivity of our results to the entropy parameterization
we adopted, we experimented with allowing an additional break at large
radius. We found that the break radius was poorly constrained 
($35^{+68}_{-23}$~kpc) and adding it did not significantly improve the fit 
($\Delta \chi^2$=3 for 2 d.o.f.). Based on the ratio of the Bayesian evidence
($6\times 10^{-3}$), the model {\em without} the break is actually preferred
at the $\sim$2.8-$\sigma$ level. Including the break had a minimal effect
on the best-fitting derived parameters (``$\Delta$Entropy'' 
in Tables~\ref{table_mass}--\ref{table_fb}).

In our default analysis, the projected temperature and density
profile were weighted by the gas emissivity,
folded through the instrumental responses
\citep[for details, see Appendix B of][]{gastaldello07a}. Since the
computation of the gas emissivity 
assumes that the three dimensional gas abundance profile
is identical to the projected profile (which is unlikely to be true), 
we explored the sensitivity of our results to this approximation
by adopting  the extreme approach of ignoring the spatial 
variation of the gas emissivity altogether. We found that this 
had a very small effect on our results
($\Delta$Weighting in Tables~\ref{table_mass}--\ref{table_fb}). 

To examine the error associated with distance uncertainties, 
we varied the distance to \src\ by $\pm$20\%, finding the effect, particularly
on \fb\ and \fgas\  to be relatively minor (``$\Delta$Distance'').
Finally, to examine the possible errors associated with our 
treatment of the covariance between the density data-points, we investigated 
adopting a more complete treatment that considers the covariance between all the 
temperature and density data-points, as well as adopting the more standard 
(but incorrect) approach of ignoring the covariance altogether. We found that
this had a non-negligible, but modest impact on the derived parameters
(``$\Delta$Covariance'' in Tables~\ref{table_mass}--\ref{table_fb}).
}

\section{Discussion} \label{sect_discussion}
\subsection{The \survey\ Galaxy Survey}
The \survey\ Galaxy Survey is intended, in part, to provide a sample
of candidate \lstar\ galaxies with hot gas halos which are likely 
to be hosted in a  galaxy-scale
(\ltsim $10^{13}$\msun) DM halo. The full sample \citep{buote12a} includes several 
objects that are known to exhibit these characteristics 
(\eg\ NGC\thin 720: \citealt{humphrey11a};  NGC\thin 4125: \citealt{humphrey06a}; NGC\thin 7796: \citealt{osullivan07b}), but the majority
of the galaxies have not been studied before in the X-ray. 
\src\ was identified as an X-ray bright object, based on the short initial
\survey\ \xmm\ observation, and targetted for deep follow-up with both
\chandra\ and \xmm. 

{The X-ray morphology is very relaxed, without obvious disturbance and there 
is no bright 
AGN (as evinced by its modest radio emission: \citealt{condon98a}, and 
the lack of a central X-ray point source), suggesting  no evidence of 
jet/ gas heating in the recent past. }
The hot gas halo is clearly detectable out to
$\sim$200~kpc ($\sim$80\%\ of \rfive, which is $240\pm22$~kpc), 
and the temperature profile is approximately 
isothermal at kT$\simeq$0.5~keV. Despite the slightly higher 
mass, the properties of \src\ are remarkably similar to NGC\thin 720, 
making \src\ one of the smallest DM halos for which interesting X-ray
constraints on the mass profile have been found, and helping to 
populate the very sparse
low-mass end of the \cvir-\mvir\ plane with vital new data. 
The properties of \src\ are, in fact, very close to our expectations for
\survey\ galaxies, providing a crucial validation of our observing strategy. 

\subsection{Hydrostatic Equilibrium} \label{sect_he}
The best-fitting hydrostatic model fits the density and temperature
data-points well, {consistent with the gas being close to hydrostatic.}
Despite nontrivial temperature and density profiles that cannot 
be parameterized by simple models individually, a smooth,
physical mass model, coupled to a monotonically rising 
entropy profile (\ie\ that is stable against convection), was able
to reproduce them well. 
If the gas is far from hydrostatic,
this would require a remarkable conspiracy between the temperature, 
density and inferred mass profiles.
The closeness of the system to hydrostatic 
is unsurprising given its relaxed X-ray morphology (\S~\ref{fig_images}).

{Nevertheless, modest deviations from hydrostatic equilibrium cannot
be entirely ruled out. 
While numerical structure formation simulations
suggest that hot halos around galaxies should, indeed, be 
quasi-hydrostatic \citep[\eg][]{crain10a}, deviations 
from the hydrostatic approximation
may introduce systematic errors on the recovered mass 
of as much as $\sim$25\%, if they are similar to clusters}
\citep[\eg][]{tsai94a,rasia06a,nagai07a,piffaretti08a,fang09a}. In order to constrain
such non-hydrostatic effects observationally, recent studies have begun to
compare  the mass distributions of galaxies inferred from 
stellar dynamics modelling to those independently obtained by 
X-rays, including several systems that are manifestly more disturbed than \src\ 
\citep[\eg][]{humphrey08a,churazov08a,churazov10a,gebhardt09a,shen10a,das10a,das11a,humphrey12b}.
Although better control of systematic errors may be necessary before 
deviations from hydrostatic equilibrium 
can be measured accurately \citep[\eg][]{churazov08a,gebhardt09a,das11a,humphrey12b},
the X-ray measurements may underestimate the mass by as much as $\sim$30\%\
in the central $\sim$5--10~kpc. In \S~\ref{sect_syserr_he}, {we explored
plausible nonthermal pressure profiles consistent with these observations,
and found that, aside from a modest increase in the total mass of the system,
if nonthermal pressure at this level is present in \src, our conclusions
about the global properties of the system were largely unaffected.}

While subject to uncertainties in the stellar initial mass
function \citep[\eg][]{treu10a,vandokkum10a} 
and the star-formation history of the galaxy, comparisons 
between the stellar M/L ratio inferred from simple stellar 
population (SSP) models and from X-ray studies 
similarly suggest that deviations from hydrostatic equilibrium
are generally small \citep{humphrey09d}. Following \citet{humphrey06a},
we estimated the stellar age ($8.3^{+2.3}_{-1.9}$~Gyr) and metallicity 
([Z/H]=$0.30^{+0.06}_{-0.08}$) of \src\ by fitting the 
models of \citet{thomas03a} to the ${\rm H\beta}$, Fe5270, Fe5335 and 
Mgb Lick indices published by \citet{ogando08a}. For a \citet{kroupa01a}
IMF and the SSP model of \citet{maraston05a}, this corresponds to 
an I-band stellar M/L ratio of 2.6$\pm$0.6\msunlsun, which agrees
well with our measurement (2.49$^{+0.26}_{-0.42}$\msunlsun). This is 
consistent with observations of other galaxies, and supports the idea
that the hydrostatic approximation is good for \src. 
{Nevertheless, given the uncertainties (both statistical and systematic) 
in this comparison, nonthermal pressure providing as much as $\sim$40\%\ 
of the total support could be consistent with the data, particularly 
if the IMF is more bottom heavy than \citeauthor{kroupa01a}; 
for a Salpeter IMF, for example, M/L=$4.0\pm0.9$, which is within
$\sim$2-$\sigma$ of our measurement.  To quantify more
precisely the accuracy of the hydrostatic approximation will likely require
sophisticated (orbit-based) stellar dynamical modelling of this triaxial
galaxy.}

\subsection{Mass profile} \label{sect_discuss_mass}
\begin{figure}[!h]
\centering
\includegraphics[width=3.4in]{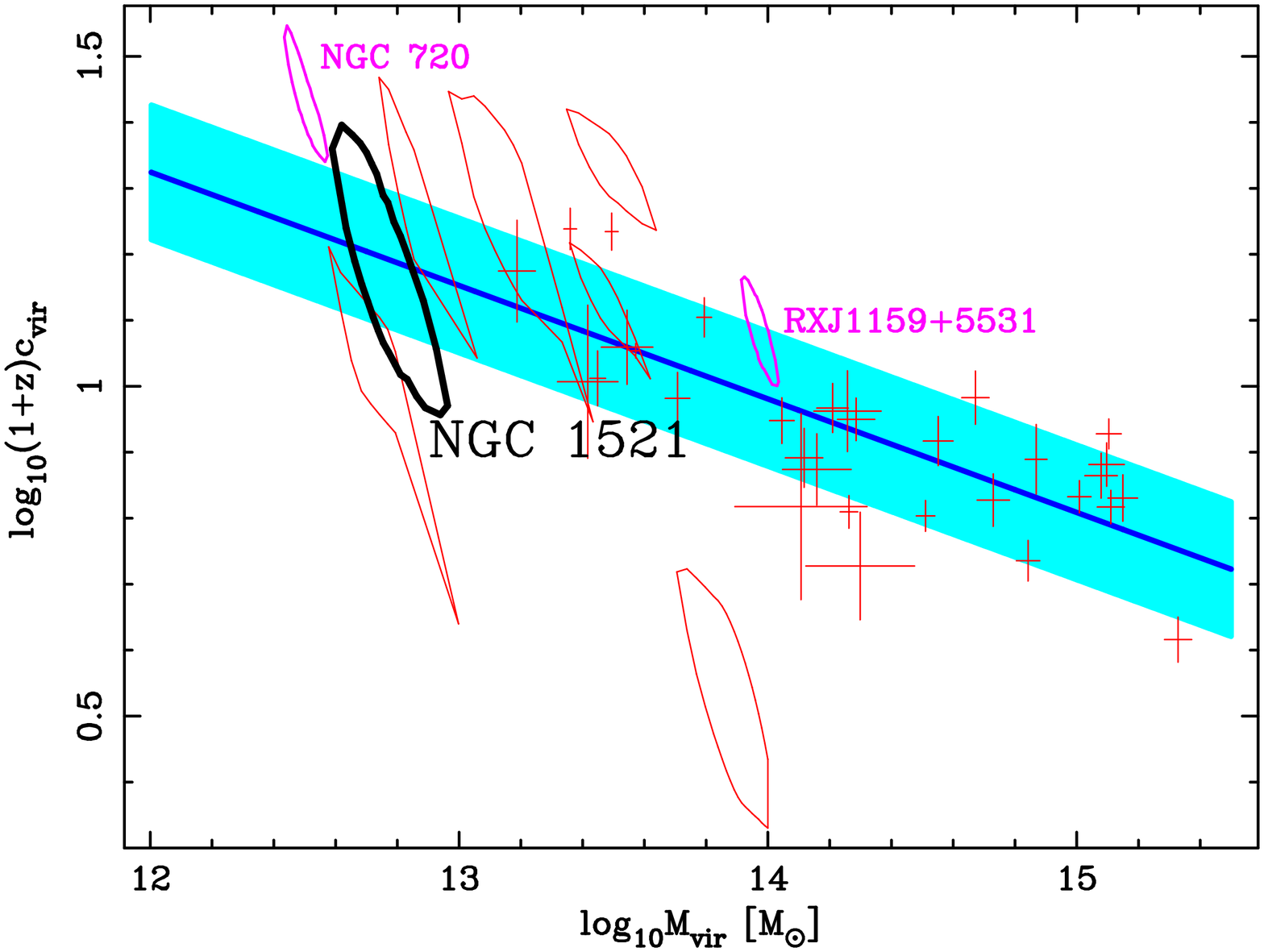}
\caption{Concentration-virial mass constraints for \src, shown along with the
sample of galaxies, groups and clusters in \citet{buote07a}. Where possible,
we show the 1-$\sigma$ confidence contours for the objects in \citet{humphrey06a},
\citet{humphrey11a} and \citet{humphrey12a} \label{fig_cm}. The solid blue line
is the best-fitting powerlaw fit obtained by \citet{buote07a}, and the 
blue shaded region indicates the 1-$\sigma$ intrinsic scatter of the points about it. 
\src\ is in good agreement with this relation.}
\end{figure}
\begin{figure}
\centering
\includegraphics[width=3.4in]{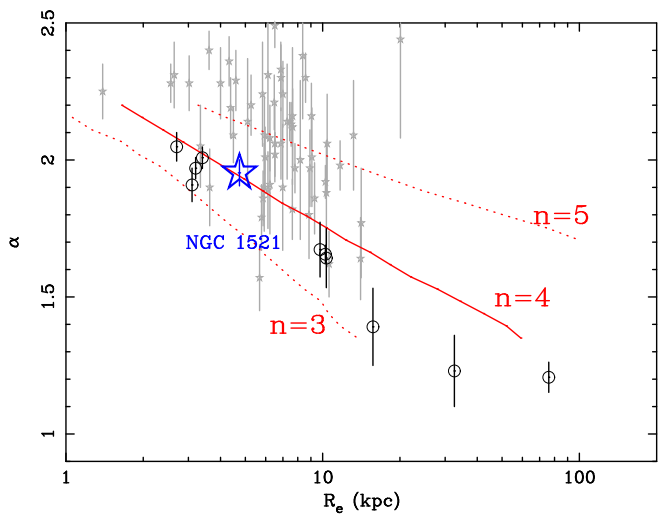}
\caption{Best-fitting (total) mass slope ($\alpha$) {\em versus} 
K-band \reff\ relation for \src\ (blue star). We overlay 
(black circles) the data from \citet{humphrey10a}, and the data derived by 
\citet{auger10a} from strong lensing and stellar kinematics within $\sim$\reff\
(grey stars). The solid and dotted (red) lines are the predictions of the toy 
model of \citet{humphrey10a} for different Sersic indices (n) for the stellar light.
\label{fig_mass_slope}}
\end{figure}
Based on our hydrostatic analysis of the \chandra\ and \xmm\
data, we were able to confirm the presence of dark matter
in \src\ at high significance (8.2-$\sigma$). The constraints
our model was able to place on the virial mass and concentration
(Fig~\ref{fig_cm})
are competitive with other X-ray determined measurements 
in this mass range \citep[\eg][]{humphrey06a}. The system is 
clearly consistent both with the empirical \citet{buote07a} 
\cvir-\mvir\ relation, and also the 
theoretical model of \citet{maccio08a} (which predicts a slightly
lower $log_{10} cvir= 0.97$ at $10^{13}$\msun). This may slightly
ease tension at the low mass end between theory and observations,
although more observations are needed to resolve this issue.
It is expected that isolated systems have higher concentrations,
on average, than the population as a whole, although the effect
is not predicted to be dramatic \citep[\eg][]{maccio08a}. Nevertheless,
\citet{khosroshahi07a} reported substantially enhanced concentrations
in a small sample of fossil groups and isolated galaxies. 
While NGC\thin 720 could be consistent with this picture, there is 
no evidence of such an effect in \src, and nor does the \survey\
galaxy NGC\thin 4125 show such a feature
\citep{humphrey06a}. A similar conclusion
was reached for the fossil groups studied by \citet{gastaldello07a}.

\src\ is a more luminous galaxy than NGC\thin 720
(by $\sim$60\%\ in the K-band, and $\sim$90\%\ in B),
and so it is not surprising that it sits in a more massive
halo (by a factor $\sim$2). The allowed mass range of \src\ 
begins to approach the regime of galaxy groups 
\citep[\gtsim $10^{13}$\msun:][]{humphrey06a},
but, since it represents one of the most luminous galaxies 
in our sample, this will likely represent the upper mass envelope
of the \survey\ galaxies. 
While the mass constraints in \src\ are competitive with the 
best measurements in \citet{humphrey06a}, we note that they 
are poorer  than for NGC\thin 720 \citep{humphrey11a}, especially
for the concentration. Although the slightly lower flux and 
shallower observations for \src\ have some role to play
in this effect, it mostly reflects degeneracies between the 
dark and luminous matter, since the dark matter only dominates
outside $\sim$20~kpc, which is comparable to the scale radius
($31^{+19}_{-11}$~kpc). Nevertheless, the virial mass is more tightly 
constrained, which reflects the good radial coverage at larger
scales.

Even though the light profile of \src\ is quite complex
(\S~\ref{sect_optical}), and so a simple interpretation
in terms of the toy Sersic+NFW model proposed by \citet{humphrey10a}
may not be possible (\ie\ the model lines in Fig~\ref{fig_mass_slope}), 
we found that 
we were able to fit the radial (total) mass distribution within 
10\reff\ with a purely powerlaw density profile distribution.
In Fig~\ref{fig_mass_slope}, we show the locus of \src\
in the $\alpha$-\reff\ plane, which 
lies very close to the $\alpha$-\reff\
relation established by other systems \citep{humphrey10a}.
This further supports the idea that approximately powerlaw total
mass distributions may be a natural consequence of the 
process of galaxy formation.



\subsection{Gas and baryon fraction} \label{sect_fb}
\begin{figure}
\centering
\includegraphics[width=3.4in]{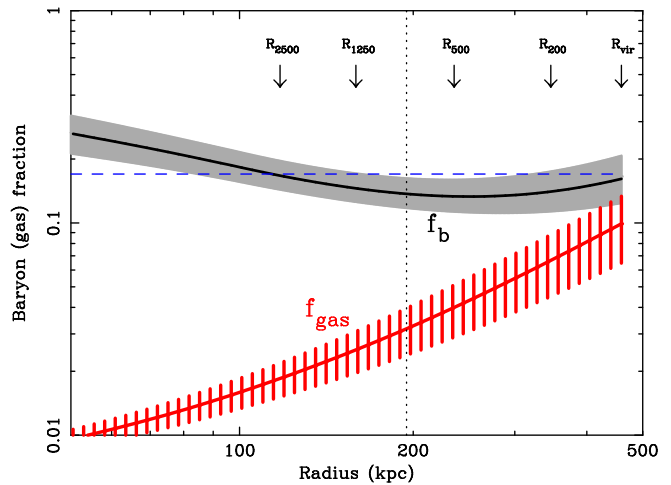}
\caption{Baryon fraction profile inferred from our best-fitting models
(black line). The shaded grey region indicates the 1-$\sigma$ statistical uncertainty
in the fits. The red shaded region indicates the gas fraction profile and 1-$\sigma$
uncertainty. The dashed line indicates the best-fitting Cosmological value of \fb, 
based on the 5-year WMAP data \citep{dunkley09a}. We indicate the physical scales
corresponding to \rvir\ and various other standard radii, and the dotted line indicates
the radial extent of the data being fitted. We note that 
the \fb\ profile is relatively flat, and is consistent with the Cosmic mean
by $\sim$\rvir. This is very similar to the profiles measured in 
NGC\thin 720 \citep{humphrey11a} and the fossil group RXJ1159+5531
\citep{humphrey12a}.  \label{fig_fb_profile}}
\end{figure}
Based on our self-consistent hydrostatic mass model, and the fit to the density
profile, we were able to constrain the gas and baryon fractions 
in \src, allowing a direct comparison with both NGC\thin 720 and more
massive groups and clusters. In Fig~\ref{fig_fb_profile}, we show the radial 
profile of \fgas\ and \fb, and we tabulate the values at various interesting
scales.
We see qualitatively the same behaviour as in 
NGC\thin 720, \ie\ an approximately flat \fb\ profile, close to the
cosmic mean. 
When evaluated at \rvir, 
\fb\ is consistent  with baryonic
closure, in agreement with the picture that the majority of the baryons
exist in the hot ISM, and the behaviour seen in NGC\thin 720 and
also the more massive fossil group RXJ1159+5531 \citep{humphrey12a}.
Within plausible systematic uncertainties, it is possible that \fbvir\
in \src\ could be slightly lower ($\sim$0.10), but even in that case
the system has retained the bulk of its baryons.
{Whether the two galaxies are representative of systems at this mass-scale, or if they 
constitute special cases, remains to be established. Results for the full \survey\ sample will help us to address this question.}


Unfortunately,  the measured value of \fb\ in \src\ does not include all
of the system's baryons. 
In particular, it ignores the 
dwarf companions, and a possible extended stellar envelope, analogous
to intracluster light. Without more velocity measurements, it is difficult 
to estimate what fraction of the galaxies within the projected 
virial radius (Fig~\ref{fig_galaxies}) should actually be included 
in such a calculation. Nevertheless, for a halo of \mvir$=6\times 10^{12}$\msun, 
anywhere between $\sim$5--40\%\ of the total stellar light could be 
in the satellite galaxies and extended envelope 
\citep{purcell07a,gonzalez07a}, which could lead to an underestimate 
in \fbvir\ of at most $\sim$0.04, which is comparable to the statistical 
errors.
We note that \src\ itself does not contain significant cool gas  
\citep{huchtmeier94a}.

In contrast, \fgas\ is more robustly known, and can easily be compared to measurements
in other systems. Within \rfive, we found that \fgas\
is in good agreement with an extrapolation
of the trends seen in galaxy groups and clusters; extrapolating the 
\fgas-\rfive\ relation from \citet{giodini09a} to the mass of \src, we 
would expect \fgas=$0.046\pm0.008$, very close to the observed value
$0.040^{+0.006}_{-0.008}$. Similar behaviour was seen in NGC\thin 720
\citep{humphrey11a}.



\subsection{Entropy profile}
\begin{figure}
\centering
\includegraphics[width=3.4in]{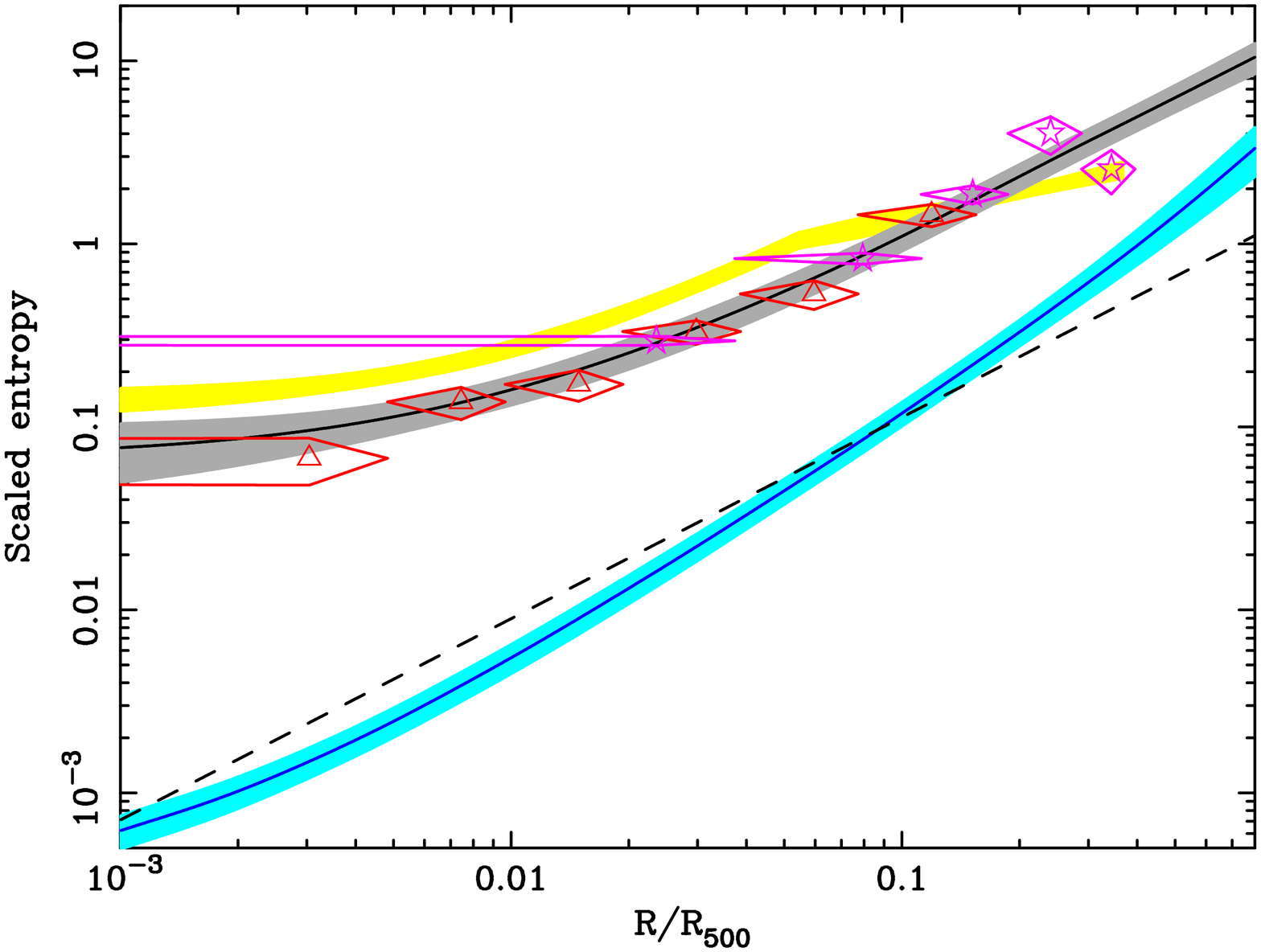}
\caption{1-$\sigma$ confidence region for the entropy profile model of \src, 
scaled by its characteristic entropy
\kfive\ (=38.5~keV~cm$^2$), and shown as a function of  \rfive. 
We overlay deprojected entropy data-points (see \S~\ref{sect_syserr_3d}),
similarly scaled. The \chandra\ data are marked with (red) triangles 
and the \xmm\ data with (magenta) stars.
The dashed line indicates the ``baseline'' prediction from gravitational
structure formation \citep{voit05b}. Also shown (blue shaded region)
is the scaled entropy model, corrected
for the gas fraction profile, which agrees better
with the baseline model. In yellow we show the entropy profile for the 
isolated galaxy NGC\thin 720 \citep{humphrey11a}. \label{fig_scaled_entropy}}
\end{figure}
As expected for approximately hydrostatic gas, we find that we obtain a good fit to the data
with a model requiring a monotonically rising entropy (S) profile. We show the model profile in
Fig~\ref{fig_scaled_entropy} (grey shaded region), scaled by the ``characteristic entropy'' $K_{500}$,
and shown as a function of fraction of \rfive\ reached. We find that the 
characteristic shape of the profile is qualitatively similar to that of NGC\thin 720, 
showing a central ``plateau'', and gradual steepening of the profile with radius, approximately reaching $S\propto R^{1.1}$, although it does not exhibit the flattening outside $\sim$0.1\rfive\
in that galaxy.
At all radii, the entropy profile is significantly enhanced over the ``baseline''
model for gravity-only structure formation simulations  \citep{voit05b}, 
indicating significant entropy injection. Given the more massive halo of \src, as compared to
NGC\thin 720, it is unsurprising that the entropy injection appears to have affected the
profile more significantly in the latter. 

To provide a less model-dependent view of the entropy profile, we overlay in Fig~\ref{fig_scaled_entropy}
a series of data-points, which are directly computed from the deprojected density and 
temperature profiles. These were obtained by emulating the behaviour of the ``projct'' \xspec\
model, and correcting for emission projected into the line of sight from outside the 
outermost annulus (see \S~\ref{sect_syserr_3d}). 
These data agree well with the smooth model,
giving us confidence in our projection procedure.

Following \citet{pratt10a}, we investigated whether scaling the entropy profile by a 
correction factor ($\left( f_g(R)/ f_{b,U}\right)^{2/3}$, where $f_g(R)$ is the gas 
fraction profile, and $f_{b,U}$ is the Cosmic baryon fraction, 0.17) 
brings it into better agreement with the baseline model.
We show this ``\fgas-corrected'' entropy profile in Fig~\ref{fig_scaled_entropy}, which clearly
agrees much better with the baseline model. This is consistent with a picture in which
entropy injection primarily manifests itself by pushing the gas out to large radii.

\subsection{Abundance profile}
An intriguing result from our study is the detection of a negative abundance gradient
in the hot ISM of \src\ (Fig~\ref{fig_abundance}), very similar to trends seen in relaxed
galaxy groups \citep[\eg][]{humphrey05a,buote02a,buote03b,buote04c,rasmussen07a}, and the 
$\sim$Milky Way-mass elliptical NGC\thin 720 \citep{humphrey11a}. This suggests that such
abundance gradients may be commonplace even in isolated (X-ray bright) elliptical galaxies, 
which has implications for the bulk gas motions responsible for distributing the metals
\citep{mathews03a}, and may imply large-scale flows driven by low-level AGN heating
\citep{mathews04a}. 

In general, the emission-weighted Fe abundance (\zfe) of the hot ISM in an early-type galaxy
is consistent with, or higher than, the metallicity of the stars \citep{humphrey05a}. 
In the case of \src, however, there appears some tension between the two measurements;
using \xmm, the Fe abundance, even averaged over the central $\sim$10~kpc is only 0.44$\pm$0.09
times Solar, while the results from fitting the published Lick indices 
\citep[see \S~\ref{sect_he};][]{ogando08a} imply a global stellar abundance of 
$1.4\pm0.3$ times Solar, 
making this system an outlier ($\sim$3.6-$\sigma$ away from equality
between the metallicity of the stars and gas) from the relation found by 
\citet{humphrey05a}. One possible solution to this problem is the ``Fe bias'', a
systematic underestimate of the Fe abundance when a single temperature plasma model is 
fitted to an inherently multi-temperature X-ray spectrum \citep{buote98c,buote00a}. 
Although the temperature profile of \src\ is close to isothermal, we experimented
with adding an additional $\sim$0.2~keV gas component to the central \xmm\ bin. 
The fit improved modestly ($\Delta$C=9.7 for 2 degrees of freedom), with $\sim$30\%\
of the gas in that bin in the cooler phase, and we found
a subtle increase in the best-fitting abundance, accompanied by a significantly 
expanded error bar (\zfe=$0.50^{+0.22}_{-0.08}$). The increase of the error bar
slightly eased tension with the other objects discussed in \citet{humphrey05a},
making it only $\sim$2.7-$\sigma$ discrepant with parity between the metallicities
of the gas and stars. We note that, although this would suggest
non-isothermal gas in the central region,  one can
still obtain reliable mass profiles with our modelling procedure by interpreting 
a single-temperature fit to the data, provided care is taken to average the model
suitably, as we did in the present work \citep{gastaldello07a}.

\acknowledgements
We would like to thank Aaron Barth and Luis Ho for kindly 
providing their optical data, taken
as part of the Carnegie-Irvine Galaxy Survey
(http://cgs.obs.carnegiescience.edu/CGS/Home.html).
We would also like to thank Fabio Gastaldello for discussions and support with the 
\xmm\ analysis. We thank Christina Topchyan for suggesting the acronym
\survey.
This research made use of the
NASA/IPAC Extragalactic Database (\ned)
which is operated by the Jet Propulsion Laboratory, California Institute of
Technology, under contract with NASA, and the HyperLEDA database
(http://leda.univ-lyon1.fr). We are grateful to the WIND-SWE team for making
their data publicly available.
PJH and DAB gratefully acknowledge support from \chandra\ award 
G09-0092X, issued by the \chandra\ X-ray Center, which is operated by 
the Smithsonian Astrophysical Observatory for and on behalf of 
NASA. Partial support for this work was also provided by NASA 
under \xmm\ grant NNX08AX74G and grant NNX10AD07G, 
issued through the office of Space Science Astrophysics Data Program. 


\bibliographystyle{apj_hyper}
\bibliography{paper_bibliography.bib}

\end{document}